\documentclass[11pt,a4paper]{emulateapj}
\usepackage{graphicx}
\usepackage{textcomp}
\usepackage{amsmath}
\usepackage{amssymb}
\usepackage{gensymb}
\usepackage{rotating}
\usepackage{epsfig}
\usepackage{longtable}
\usepackage{color}
 \def\lha{$L_{{\rm H}\alpha}$}
 \def\lhamath{L_{{\rm H}\alpha}}

 \newcommand{\aips}{{$\cal AIPS\/$}}

 \def\herschel{{\it Herschel}}
 \def\spitzer{{\it Spitzer}}

 \def\l14{$L_{\rm 1.4GHz}$}
 \def\s14{$S_{\rm 1.4GHz}$}

 \def\jthreetwo{$J\!=\!3\!-\!2$}

 \newcommand{\co}{$^{12}$CO}
 \newcommand{\etal}{et~al.}
 \newcommand{\hst}{\textit{HST}}

 \def\gs{\mathrel{\raise0.35ex\hbox{$\scriptstyle >$}\kern-0.6em\lower0.40ex\hbox{{$\scriptstyle \sim$}}}}
 \def\ls{\mathrel{\raise0.35ex\hbox{$\scriptstyle <$}\kern-0.6em\lower0.40ex\hbox{{$\scriptstyle \sim$}}}}

 \def\Wm2{\,\hbox{W}\,\hbox{m}^{-2}}
 \def\gsim{\mathrel{\raise0.35ex\hbox{$\scriptstyle >$}\kern-0.6em\lower0.40ex\hbox{{$\scriptstyle \sim$}}}}
 \def\lsim{\mathrel{\raise0.35ex\hbox{$\scriptstyle <$}\kern-0.6em\lower0.40ex\hbox{{$\scriptstyle \sim$}}}}

\lefthead{Thomson et al.}  \righthead{The dust and gas properties of stacked H$\alpha$-selected galaxies}

\begin{document}

\title{Evolution of dust-obscured star formation and gas to $z=2.2$ from HiZELS}

\author{
A.\,P.\ Thomson,\altaffilmark{1}
J.\,M.\ Simpson,\altaffilmark{2}
Ian Smail,\altaffilmark{1}
A.\,M.\ Swinbank,\altaffilmark{1}
P.\,N.\ Best,\altaffilmark{2}
D.\ Sobral,\altaffilmark{3}
J.\,E.\ Geach,\altaffilmark{4}
E.\ Ibar,\altaffilmark{5}
H.\,L.\ Johnson,\altaffilmark{1}
}

\setcounter{footnote}{0}
\altaffiltext{1}{Centre for Extragalactic Astronomy, Department of Physics, Durham University, South Road, Durham DH1 3LE, UK; email:  alasdair.thomson@durham.ac.uk}
\altaffiltext{2}{Scottish Universities Physics Alliance, Institute for Astronomy, Royal Observatory Edinburgh, Blackford Hill, Edinburgh, EH9 3HJ, UK}
\altaffiltext{3}{Department of Physics, Lancaster University, Lancaster, LA1 4YB, UK}
\altaffiltext{4}{Centre for Astrophysics Research, Science \& Technology Research Institute, University of Hertfordshire, Hatfield, AL10 9AB, UK}
\altaffiltext{5}{Instituto de F\'isica y Astronom\'ia, Universidad de Valpara\'iso, Avda. Gran Breta\~na 1111, Valpara\'iso, Chile}

\begin{abstract} We investigate the far-infrared (far-IR) properties of galaxies selected via deep, narrow-band imaging of the H$\alpha$ emission line in four redshift slices from $z=0.40$--$2.23$ over $\sim 1$\,deg$^2$ as part of the High-redshift Emission Line Survey (HiZELS). We use a stacking approach in the \herschel\ PACS/SPIRE far-IR bands, along with $850\,\mu$m imaging from SCUBA-2 and Very Large Array (VLA) 1.4\,GHz imaging to study the evolution of the dust properties of H$\alpha$-emitters selected above an evolving characteristic luminosity threshold, $0.2L^\star_{{\rm H}\alpha}(z)$. We investigate the relationship between the dust temperatures, $T_{\rm dust}$, and the far-infrared luminosities, $L_{\rm IR}$ of our stacked samples, finding that our H$\alpha$-selection identifies cold, low-$L_{\rm IR}$ galaxies ($T_{\rm dust}\sim 14$\,{\sc k}; $\log[L_{\rm IR}/{\rm L}_\odot]\sim 9.9$) at $z=0.40$, and more luminous, warmer systems ($T_{\rm dust}\sim 34$\,{\sc k}; $\log[L_{\rm IR}/{\rm L}_\odot]\sim 11.5$) at $z=2.23$. Using a modified grey-body model, we estimate ``characteristic sizes'' for the dust-emitting regions of H$\alpha$-selected galaxies of $\sim 0.5$\,kpc, nearly an order of magnitude smaller than their stellar continuum sizes, which may provide indirect evidence of clumpy ISM structure. Lastly, we use measurements of the dust masses from our far-IR stacking along with metallicity-dependent gas-to-dust ratios ($\delta_{\rm GDR}$) to measure typical molecular gas masses of $\sim 1\times 10^{10}$M$_\odot$ for these bright H$\alpha$-emitters. The gas depletion timescales are shorter than the Hubble time at each redshift, suggesting probable replenishment of their gas reservoirs from the intergalactic medium. Based on the number density of H$\alpha$-selected galaxies, we find that typical star-forming galaxies brighter than $0.2L^{\star}_{{\rm H}\alpha}(z)$ comprise a significant fraction ($35\pm10$\%) of the total gas content of the Universe, consistent with the predictions of the latest state-of-the-art cosmological simulations.
\end{abstract}


\section{Introduction}\label{sec:intro}

It is now widely established that the star-formation rate density (SFRD) of the Universe increased from early times, reaching a peak around $z\sim1$--$2$, and has been in steady decline ever since \citep[e.g.\,][]{lilly96, hopkins06, madau14}. Similar evolution been claimed for the typical ratio of the star-formation rate (SFR) and stellar mass ($M_{\star}$), the specific star-formation rate \citep[sSFR, e.g.\,][]{magdis10}. This has been linked, in part, to changes in the availability of molecular gas (the raw fuel for star formation) as a function of redshift \citep[e.g.\,][]{genzel10}, and also potentially results in changes in the morphologies of star-forming galaxies, from smooth, disk-like structures in the local Universe to more clumpy, irregular morphologies seen at $z\sim2$ \citep[e.g.\,][]{elmegreen09, swinbank12a}.

However, existing studies of the evolving properties of star-forming galaxies are based on a range of techniques both for selecting star-forming galaxies, and for measuring their ``instantaneous'' SFRs \citep[see e.g.\,][]{kennicutt12}. These include galaxies selected: via their Ly$\alpha$\ line emission \citep[LAEs;][]{partridge67}, which are predominantly blue, less-massive galaxies \citep{oteo15}; via the ``break'' in their spectra near the Lyman limit at $912$\,{\mbox{\normalfont\AA}} \citep[Lyman Break Galaxies; ][]{steidel96}; through their rest-frame UV-optical colours \citep[e.g.\,$BzK$;][]{daddi04}; and galaxies selected via their rest-frame far-infrared (IR) emission, which traces the dust heated by intense star formation episodes, such as the Ultra-Luminous InfraRed Galaxies \citep[e.g.\,][]{lonsdale90} and their high-redshift analogues, the sub-millimetre galaxies \citep[SMGs;][]{smail97, barger98, casey14}.  The varying utility of these different techniques, each with their own selection functions with redshift, SFR and extinction has complicated the process of establishing a unified view of the cosmic star formation history. 

Overcoming these limitations requires the use of homogeneous samples of star-forming galaxies selected from large-area, clean and deep multiwavelength observations, which both minimise cosmic variance, and circumvent the need to perform extrapolations down to faint luminosities. Narrow-band imaging techniques provide one means of making progress, as they can be used in sensitive, wide-field surveys to select large, representative samples of ``typical'' star-forming galaxies in a relatively clean way from a single emission line, which can act as an indicator for the SFR from low to high redshift. The H$\alpha$ line ($\lambda_{\rm rest} = 656.3$\,nm) is one of the most commonly-employed SFR indicators, and has been used widely for star-forming galaxies at $z\lesssim 3$ \citep[e.g.\,][]{kewley02,sobral13}. H$\alpha$ emission arises predominantly from young, massive OB stars ($\lesssim 10$\,Myr old, and $\gtrsim 8$\,M$_{\odot}$), and can be used to measure the SFR, if corrected for the extinction ($A_{{\rm H}\alpha}$) due to scattering and absorption by dust \citep{kennicutt98}. 

Recent work in the optical/near-IR has revealed the existence of an apparent correlation between the SFR and stellar mass ($M_{\star}$), the so-called ``main sequence'', whose logarithmic slope is thought to be close to linear \citep{elbaz07, daddi07}, with evidence of a possible deviation to sub-linear slopes \citep[$\sim0.6$--$0.8$; ][]{whitaker12, magnelli14} at high $M_{\star}$. Recently, \citet{schreiber16} presented an analysis of the far-IR properties of a mass-selected sample of main sequence galaxies ($M_\star \gtrsim 10^{10}$\,M$_\odot$), using a stacking analysis to peer below the confusion limit of deep 250, 350 and 500\,$\mu$m images, taken as part of the GOODS-\herschel\footnote{\herschel\ was an ESA space observatory with science instruments provided by European-led Principal Investigator consortia and with important participation from NASA.} and CANDELS-\herschel\ key programs. They explain the flattening of the main sequence at high-$M_{\star}$ not as being due to high-$M_\star$ galaxies lacking fuel (i.e.\,low $M_{\rm gas}$), but rather due to their having a lower star-formation efficiency (${\rm SFE}\equiv{\rm SFR}/M_{\rm gas}$). Possible explanations for this low-SFE are that the SFR is artificially suppressed, either (i) by radio-mode AGN feedback \citep[e.g.\,][]{bower06}, which drives gas out of a galaxy, preventing it from cooling to form stars, or (ii) by morphological quenching \citep[e.g.\,][]{martig09}, wherein the internal kinematics of a galaxy's stellar disk set up differential torques, which act to prevent the cold gas clouds from fragmenting. An alternative explanation for flattening of the main sequence in high-$M_{\star}$ galaxies is simply that the most massive galaxies contain old stellar bulges which, are almost completely decoupled from the ongoing star formation (and the molecular gas reservoir that fuels it), and that by instead considering the stellar mass only in the disk, a constant main sequence slope emerges \citep{abramson14}.

To fully understand the evolution of star-forming galaxies, we must also investigate the link between star formation and its fuel supply, by tracking the typical molecular gas mass, $M_{\rm H_2}$, at different epochs. However, this is observationally challenging; H$_2$ lacks a strong dipole and thus does not radiate strongly \citep{carilli13}. Tracer molecules such as \co, the second most abundant molecule in the interstellar medium (ISM), have been observed out to $z\sim 6$ \citep[e.g.\,][]{riechers13}, but are observationally expensive to detect, requiring long integrations with interferometers, hence it is difficult to construct statistical samples \citep[e.g.\,][]{thomson12, bothwell13}. However, some progress can be made by using the relationship between dust and gas in local galaxies, which allows estimates of the gas mass to be obtained very quickly via observations of the Rayleigh-Jeans continuum \citep[see][for a thorough discussion of this method]{scoville14}.

Here we combine results from a narrow-band H$\alpha$ survey with measurements of the dust content of H$\alpha$-selected galaxies in order to address the issue of the evolution of normal star-forming galaxies. 

The High-redshift(Z) Emission Line Survey \citep[HiZELS: ][]{geach08} conducted observations through specially-designed, narrow-band filters on the Wide Field CAMera (WFCAM) on the United Kingdom Infra-Red Telescope  (UKIRT\footnote{UKIRT programmes U/CMP/3 and U/10B/07}). HiZELS used narrow-band filters in the $J$, $H$ and $K$ bands (NB$_{\rm J}$, NB$_{\rm H}$ and NB$_{\rm K}$), corresponding to redshifted H$\alpha$\ emission at $z=0.84, 1.47$ and $2.23$, along with complementary observations with the Subaru telescope in the NB921 filter\footnote{Subaru programme S10B-144S} \citep{sobral12}, which are sensitive to  H$\alpha$\ emission at $z=0.40$.

Prior to the launch of \textit{Herschel}, far-IR studies of (SFR-selected) HiZELS galaxies by \citet{geach08} and \citet{garn10} relied on \spitzer\ 24, 70 and 160\,$\mu$m imaging -- in many cases, offering only upper-limits -- to constrain the dust properties in two redshift slices at $z=0.84$ and $z=2.23$. Later, \citet{ibar13} used \herschel\ PACS/SPIRE data covering the peak of the dust spectral energy distribution (SED) to investigate the far-IR properties of the HiZELS sample at $z=1.47$, finding that H$\alpha$-selected galaxies are highly-efficient star-forming systems, which lie somewhat above the Main Sequence. 

In this paper we build upon this earlier work, by investigating the relationships between SFR, $M_{\star}$, H$\alpha$ extinction ($A_{{\rm H}\alpha}$), dust mass ($M_{\rm dust}$) and temperature ($T_{\rm dust}$) across all four redshifts surveyed by HiZELS, in a self-consistent manner. We use the deepest available wide area images of the COSMOS and UDS fields in five far-IR bands covering the dust peak ($100$--$500$\,$\mu$m), and supplement this with new photometry at $850\,\mu$m from the Submillimeter Common-User Bolometer Array 2 (SCUBA-2) to trace the dust mass, as well as $1.4$\,GHz radio imaging from the Karl\,G.\ Jansky Very Large Array (VLA). We employ stacking techniques to circumvent the limitations created by the poor angular resolution of these images, allowing us to study the dust properties of moderately star-forming galaxies (${\rm SFR}\sim 20$\,M$_{\odot}$\,yr$^{-1}$) selected in a uniform manner across the full redshift range of the HiZELS survey.

This paper is laid out as follows:  in \S\,\ref{sec:data-reduction}, we present our analysis, including an outline of the sample selection, a description of the method used to correct the H$\alpha$ luminosities of our sample for extinction, and details of our stacking analysis. In \S\,\ref{sec:results} we discuss our main results, beginning with the luminosities, dust masses and temperatures derived from our SED fits, and a comparison of the IR and H$\alpha$-derived SFRs, which offers additional insight in to the extinction of H$\alpha$ emission by dust. In \S\,\ref{sec:madau}, we develop a framework in which to use the ``fundamental metallicity relation'' to constrain the gas-to-dust ratios of H$\alpha$-selected galaxies, and so estimate the contribution made by H$\alpha$-selected star-forming galaxies to the total H$_2$ content of the Universe. We give our main conclusions in \S\,\ref{sec:conclusions}. Throughout, we use \textit{Planck} cosmology, with ${\rm H}_0=70$\,km\,s$^{-1}$\,Mpc$^{-1}$, $k=0$, $\Omega_m = 0.3$, and $\Omega_\lambda = 0.7$, and assume a \citet{chabrier03} initial mass function (IMF).

\newpage
\section{Analysis}\label{sec:data-reduction}
\subsection{Sample selection and observations}\label{sec:sample}

The starting point of our analysis is the catalogue of 3004 H$\alpha$-selected sources from HiZELS presented in \citet{sobral13}, comprising 1771 and 1233 H$\alpha$-selected star-forming galaxies in the COSMOS and UDS fields, respectively. These total 1108 galaxies at $z=0.40$, 635 at $z=0.84$, 511 at $z=1.47$ and 750 at $z=2.23$, down to typical SFR limits of $\sim 0.1, 1.0, 2.0$ and $4$\,M$_{\odot}$\,yr$^{-1}$, respectively. We summarise the properties of the sample in Table\,\ref{tab:stack-properties}.

\begin{table*}
\centering
\caption[Properties of the HiZELS sample]{Properties of the HiZELS sample}
\label{tab:stack-properties}
\begin{tabular}{lrcccccccc}
\hline
\multicolumn{1}{c}{} &
\multicolumn{1}{r}{} &
\multicolumn{1}{c}{$N_{\rm galaxies}^a$} &
\multicolumn{1}{c}{Volume} &
\multicolumn{1}{c}{SFR limit$^b$} & 
\multicolumn{1}{c}{$N_{\rm stack}^c$} &
\multicolumn{1}{c}{$\log_{10}\bigl(L^{\star}({\rm H}\alpha)\bigr)$} & 
\multicolumn{1}{c}{$\langle {\rm SFR}_{{\rm H}\alpha}\rangle^d$} & 
\multicolumn{1}{c}{$\langle M_\star^{{\rm H}\alpha} \rangle^e$} & \\
\multicolumn{1}{c}{} &
\multicolumn{1}{r}{} &
\multicolumn{1}{c}{} &
\multicolumn{1}{c}{($10^4$\,Mpc$^3$\,deg$^{-2}$)} &
\multicolumn{1}{c}{(${\rm M}_{{\odot}}$\,yr$^{-1}$)} &
\multicolumn{1}{c}{} & 
\multicolumn{1}{c}{(erg\,s$^{-1}$)} & 
\multicolumn{1}{c}{$({\rm M}_{\odot}\,{\rm yr}^{-1})$} & 
\multicolumn{1}{c}{$(\times 10^{10}{\rm M}_{\odot})$} &\\
\hline
$z=0.40$      &          & 1108& 5.13  & --    & 52 & $42.15^{+0.47}_{-0.12}$ & 2.1  & $3.2\pm 0.9$   &\\
              &COSMOS    & 445 & --    & $0.1$ & 36 & --                   & --   & --   &\\
              &UDS       & 663 & --    & $0.2$ & 16 & --                   & --   & --   &\\
\hline
$z=0.84$      &          & 635 & 14.65 & --    & 397& $42.37^{+0.07}_{-0.05}$ & 6.9  & $0.9\pm 0.2$   &\\
              &COSMOS    & 425 & --    & $0.9$ & 240& --                   & --   & --   &\\
              &UDS       & 210 & --    & $0.9$ & 152& --                   & --   & --   &\\
\hline
$z=1.47$      &          & 511  & 33.96 & --   & 449& $42.75^{+0.06}_{-0.05}$ & 26.0 & $1.2\pm 0.3$  &\\
              &COSMOS    & 323 & --    & $1.9$ & 274& --                   & --   & --   &\\
              &UDS       & 188 & --    & $4.0$ & 175& --                   & --   & --   &\\
\hline
$z=2.23$      &          & 750 & 38.31 & --    & 535& $43.17^{+0.08}_{-0.06}$ & 34.4 & $2.3\pm 0.4$  &\\
              &COSMOS    & 578 & --    & $3.5$ & 388& --                   & --   & --   &\\
              &UDS       & 172 & --    & $7.7$ & 146& --                   & --   & --   &\\
\hline
\end{tabular}
{\small

Notes: $^a$Total number of H$\alpha$-selected star-forming galaxies in each field, at each redshift. $^b$SFR limit determined by converting the faintest H$\alpha$ luminosity in each sub-sample, using the \citet{kennicutt98} conversion factor. $^c$ Number of galaxies in each stack, selected with $L_{{\rm H}\alpha}\geq 0.2L^{\star}_{{\rm H}\alpha}(z)$. $^{d}$ Median dust-corrected H$\alpha$-derived SFR of galaxies contributing to stacked sub-samples. $^e$ Representative H$\alpha$-weighted stellar mass of each sub-sample (See \S\,\ref{sect:mstar}).
}
\end{table*}

We exploit the multitude of mid/far-IR data available in the UDS and COSMOS fields, comprising: (i) $250$, $350$ and $500$\,$\mu$m \herschel\ SPIRE observations (see \S\,\ref{sect:spire-deblending}); and (ii) cold dust-sensitive 850\,$\mu$m observations taken with SCUBA-2 on the James Clerk Maxwell Telescope (JCMT) as part of the SCUBA-2 Cosmology Legacy Survey \citep[S2CLS:][]{geach13,geach16}. In addition, in the COSMOS field, we include 100 and 160\,$\mu$m \herschel\ PACS observations from the PACS Evolutionary Probe \citep[PEP;][]{lutz11} survey, and 1.4\,GHz radio continuum observations from the Very Large Array, taken as part of the VLA-COSMOS survey \citep{schinnerer04}. 

To study evolution in the properties of ``normal'' star-forming galaxies, we select galaxies close to the knee of the H$\alpha$\ luminosity function, with dust-corrected luminosities $\lhamath \geq 0.2L^{\star}_{{\rm H}\alpha}(z)$ \citep[where $L^{\star}_{{\rm H}\alpha}(z)$ denotes the characteristic ``break'' in the H$\alpha$ luminosity function at each HiZELS redshift, measured by][]{sobral14}.

Dividing the HiZELS population on an evolving luminosity cut in this manner gives us four matched, SFR-selected sub-samples, comprising typical star-forming galaxies spanning the characteristic H$\alpha$ luminosity at each redshift. 

\subsection{Dust corrections}\label{sec:dust-corrections}
In order to account for obscuration of H$\alpha$ photons by dust within the ISM of each HiZELS galaxy, we perform an empirical dust-correction to the measured line fluxes. We begin with the observed H$\alpha$ luminosities ($L_{{\rm H}\alpha,{\rm obs}}$) and stellar continuum reddening values, $E(B-V)$ for each galaxy, reported by \citet{sobral14}. Next, we use the reddening law of \citet{calzetti00} to estimate the H$\alpha$ extinction arising from diffuse dust, $A_{{\rm H}\alpha, {\rm cont}}$ = $\kappa(\lambda)A_{V}/R'_{V}$, where $A_V = 3\times E(B-V)$ is the stellar extinction. For $\lambda_{{\rm H}\alpha, {\rm rest}}=656$\,nm, $\kappa(\lambda)$ is expressed as

\begin{center}
\begin{equation}
\kappa(656\,{\rm nm}) = 2.659\biggl(-1.857 + \frac{1.040}{\lambda / \mu{\rm m}}\biggr) + R'_{V}
\end{equation}
\end{center}

$R'_{V}=4.08\pm0.88$ is the typical effective obscuration at $V$-band \citep{calzetti00}. In addition to the extinction from the diffuse ISM, observations of nearby starburst galaxies by \citet{calzetti94} and \citet{calzetti00} indicate the need for an ``extra'' extinction component, $A_{{\rm H}\alpha, {\rm ext}}$, to account for attenuation occurring locally in the birth clouds around young OB stars. Using deep, multi-wavelength \textit{Hubble Space Telescope} (\hst) data from the CANDELS project, \citet{wuyts13} propose an empirical method to estimate $A_{{\rm H}\alpha{\rm , extra}}$, which is based in part on the dependency of the average molecular cloud mass on the galaxy-integrated gas fraction: 

\begin{center}
\begin{equation}\label{eq:wuyts}
A_{{\rm H}\alpha, {\rm ext}} = A_{{\rm H}\alpha, {\rm cont}}\bigl(0.90-0.15A_{{\rm H}\alpha, {\rm cont}}\bigr)
\end{equation}
\end{center} 

We hence determine the intrinsic (i.e.\,dust-corrected) H$\alpha$ luminosities of our H$\alpha$-selected galaxies, $L_{{\rm H}\alpha, {\rm int}}$ (hereafter, $L_{{\rm H}\alpha}$) from their uncorrected luminosities ($L_{{\rm H}\alpha, {\rm obs}}$) using:

\begin{center}
\begin{equation}
\log_{10}\bigl(L_{{\rm H}\alpha,{\rm int}}\bigr) = \log_{10}\bigl(L_{{\rm H}\alpha, {\rm obs}}\bigr) + 0.4A_{{\rm H}\alpha, {\rm cont}} + 0.4A_{{\rm H}\alpha, {\rm ext}}
\end{equation}
\end{center}

\subsection{\herschel\ SPIRE deblending}\label{sect:spire-deblending}

The \textit{Herschel} SPIRE mosaics of the UDS and COSMOS fields were taken as part of the HerMES survey \citep{oliver12}, and were retrieved from the DR2 HerMES data release. Due to the coarse resolution of the SPIRE maps (15, 22 and 32$''$ at 250, 350 and 500\,$\mu$m, respectively), the images need to be deblended before reliable flux densities can be derived. This deblending allows us to overcome the confusion limit, which smears together nearby sources, and can bias flux density measurements -- even in stacks \citep[e.g.\,][]{oliver12, magnelli13}. 

To create the prior catalogue, we used the \textit{Spitzer}\,/\,MIPS 24$\mu$m photometric catalogues from \citet{magnelli13} which are derived by simultaneous point spread function fitting to the prior positions at IRAC\,/\,3.6\,$\mu$m.  The 24\,$\mu$m catalogues were limited to a 3$\sigma$ detection limit of 50\,$\mu$Jy. To deblend the images, we followed the same procedure described in \citet{swinbank14} \citep[see also][]{stanley15}. For each of the SPIRE bands, we created a model image, in which flux was added at the positions of the galaxies in the prior catalogue and convolved with the SPIRE beam, before a residual map ($data - model$) was created. The fluxes of the galaxies in the model image were then randomly perturbed, and the process repeated until the residual map converged on a minimum. Since $\sim10$\% of the H$\alpha$-selected galaxies in HiZELS are also 24\,$\mu$m sources, their flux has been removed from the image when creating the residual image during the deblending process and so this flux needs to be added back in to the map before stacking. We therefore reinsert the flux of each 24\,$\mu$m detected HiZELS galaxy back in to each of the SPIRE residual images using the appropriate PSF.  This finally leaves us with a deblended, residual image (that includes the HiZELS galaxies that were also 24\,$\mu$m sources) that can be stacked to investigate the far-infrared properties of our sample. The decision to use 24$\mu$m/radio-detected galaxies as priors for the deblending is motivated by the assumption that those galaxies that dominate the mid-IR/radio bands will account for the majority of the total flux in the SPIRE images. However, the inherent clustering of field galaxies which are \textit{not} in the prior catalogue (but do still contribute flux to the SPIRE images) could bias the results of our stacks, if not carefully accounted for. Due to the increasing beam size at longer wavelengths, we may expect any such bias to be more severe at $500\mu$m than at shorter wavelengths. We performed a series of tests (Appendix\,A\ref{app:clustering}) to quantify this effect, but found the bias in our stacked flux densities due to clustering to be comparable to (or smaller than) the statistical uncertainties on the photometry. Moreover the ``bias'' was not found to be dependent on wavelength; we therefore opted not to apply a systematic correction to our flux densities after stacking in the residual images.

The smaller beam sizes in the PACS 100/160\,$\mu$m, SCUBA-2 870\,$\mu$m and VLA 1.4\,GHz images, coupled with the low surface density of 870\,$\mu$m sources \citep{chen16} mean that the images at these wavelengths do not require deblending. Hence, we stack at the positions of HiZELS galaxies directly in the calibrated PACS, SCUBA-2 and VLA maps, and in the residual, deblended SPIRE images.

\subsection{AGN contribution}

While the H$\alpha$ line is commonly used as a SFR indicator, the intense UV radiation fields that are responsible for its production can also be found in the vicinity of active galactic nuclei (AGN); therefore it is prudent to consider the contribution made by non star formation-dominated H$\alpha$-selected galaxies to the stacked far-IR flux densities. We identify 41 H$\alpha$-emitting AGN by searching for X-ray sources within $1''$ of each HiZELS galaxy in the \textit{Chandra} COSMOS Legacy Survey \citep{civano16} and Subaru-\textit{XMM} Deep Survey \citep[SXDS][]{ueda08} catalogues. Of these 41 AGN, none fall within the $L_{{\rm H}\alpha}\geq 0.2L^{\star}_{{\rm H}\alpha}$ sample at $z=0.40$, however there is 1 X-ray AGN in the sample at $z=0.84$, and there are 11 (2.4\% of the sample) at $z=1.47$ and 7 (1.2\%) at $z=2.23$. We note that of the 41 X-Ray AGN, only 16 are individually detected in any far-IR/radio band \citep[see also][]{calhau17}. In addition to the 41 X-ray AGN, we identify a further 17 candidate AGN via their mid-IR colours (0, 5, 5 and 7 at the respective HiZELS redshifts), following the technique of \citet{donley12}.

We performed our stacking analysis both with these sources present and with them excluded, finding the fluxes in both cases to be consistent within the errors. In an analysis which involved stacking all the candidate AGN in HiZELS at $z=1.47$, \citet{ibar13} found the far-IR SEDs of AGN to be potentially more luminous ($\log[L_{\rm IR}/{\rm L}_{\odot}]\sim 11.6$) and warmer ($\Delta T_{\rm dust}\sim 7$\,{\sc k}) on average than purely star-forming galaxies, but with significant uncertainties due to the small sample size. Given the small number of X-ray or IRAC-identified AGN in our $>0.2L^{\star}_{{\rm H}\alpha}(z)$ sub-samples, the lack of any measurable effect on the SEDs regardless of whether or not we exclude them from our stacks is therefore unsurprising. Nevertheless, we exclude all 58 candidate AGN from our subsequent analysis.

\subsection{Stacking analysis}\label{sec:stacking}

We use the deepest available extragalactic far-IR/sub-mm observations from \herschel\ and SCUBA-2. Focusing our study only on those H$\alpha$-selected galaxies that are individually detected in the far-IR/sub-mm data would introduce a strong bias towards galaxies with the most extreme SFRs, characteristic of ``starburst'' systems; in order to study \textit{typical} star-forming galaxies, and understand their dust properties, it is necessary to peer below the confusion limit in the far-IR wave-bands, and use a stacking approach.

Given the similar depths of the original H$\alpha$ observations from which the COSMOS and UDS samples were drawn ($\log_{10}\bigl[\lhamath/{\rm erg}\,{\rm s}^{-1}\bigr]\geq 39.95$), we opt to simultaneously stack in both fields.

We begin by separating the COSMOS and UDS galaxies, extracting thumbnails around each galaxy above our $0.2L^{\star}_{{\rm H}\alpha}(z)$ cuts, and running a median stacking algorithm at each wavelength on these thumbnails. In each of the far-IR/sub-mm maps, we account for the background emission by subtracting the median flux of 1000 random positions from the map. In the PACS 100 and 160\,$\mu$m bands, we measure fluxes by summing the pixel values of the stacks within a $7.2''$ or $12''$ aperture, respectively, and multiplying by the empirically-derived aperture correction ($a(\lambda)$) and high-pass filtering correction ($g(\lambda)$) specified in the PACS PEP data release notes\footnote{http://www.mpe.mpg.de/resources/PEP/DR1\_tarballs/\\readme\_PEP\_global.pdf}. 

In the de-blended SPIRE $250$, $350$ and $500$\,$\mu$m and SCUBA-2 $850$\,$\mu$m bands, the angular resolution is low enough ($\geq 15''$) that we do not resolve the emission; hence we measure fluxes from the peak pixel value within $1''$ of the centroid of each stack (to allow for any small systematic misalignments in the astrometry of the images). 

At the native $1.5''$ resolution of the VLA COSMOS image, we anticipate that some of our sources may indeed be resolved -- because we are only interested in measuring the total fluxes of the stacks, we convolve the VLA image with a $5''$ Gaussian kernel prior to stacking, using the Astronomical Image Processing System (\aips) task {\sc convl} (setting the {\sc factor} parameter to scale the map by the ratio of beam areas, in order to preserve the absolute flux scale). We then measure the $1.4$\,GHz flux densities of each of the stacked, smoothed radio images by measuring the peak pixel value within $1''$ of the centroid of the stack.

In addition to creating the median stacks at each wavelength, we also create error images by bootstrap re-sampling each of the thumbnails that are used in the stacking procedure. We determine the flux uncertainties in our stacks from these error images in the same manner as we measured the fluxes from the corresponding stacks, i.e.\,aperture photometry in the PACS images, and taking peak pixel values in the SPIRE, SCUBA-2 and VLA images. The derived flux densities are reported in Table\,\ref{tab:hizels-photometry}.

\subsection{SED fitting}\label{sec:sedfitting}
We create a set of composite photometry at each redshift, comprising the combined flux densities at $250$, $350$, $500$ and $850$\,$\mu$m from COSMOS and UDS (weighted by the number of sources in each field), and the 100 and 160\,$\mu$m, plus 1.4\,GHz fluxes from the COSMOS-only stacks (since we lack coverage in these bands in the UDS field). Combining the photometry in this manner has the effect of reducing the flux uncertainties near the peak of the SED.

We measure the properties of our galaxy stacks from this set of photometry by fitting isothermal modified blackbody (greybody) templates\footnote{We also try fitting dual-temperature greybody SEDs, and check for improvements in the fit by measuring the Bayesian Information Criterion parameter, ${\rm BIC}\equiv \chi^2 + k\ln\,N$, where $k$ is the number of degrees of freedom of the model and $N$ is the number of data points constraining the fit \citep{wit12}. BIC favours models which improve $\chi^2$, but penalises models which require several extra degrees of freedom to deliver marginal improvements in $\chi^2$.  We find in three out of four cases that the BIC formally favours an isothermal fit, however at $z=0.84$, the two-component fit lowers BIC by $\sim 50$\%. The total luminosity and dust mass of the $z=0.84$ two-component fit are $\log_{10}(L_{\rm IR}/{\rm L}_{\odot})=(10.95\pm0.15)$ and $M_{\rm dust}=(1.0\pm 0.8)\times 10^{8}$\,M$_{\odot}$, respectively, and the luminosity-weighted temperature is $T_{\rm dust}=35\pm 7$\,{\sc k}. In the interest of consistency across all four redshifts, we hereafter use properties measured from the single-temperature dust SEDs, but incorporate the differences between properties measured from the isothermal and dual-temperature fits into the errors on all derived quantities in Table\,\ref{tab:hizels-properties} as a model-dependent systematic uncertainty.} of the form

\begin{equation}\label{eq:greybody}
S(\nu)\propto\frac{\nu^{\beta+3}}{\exp{(h\nu/kT_{\rm d})}-1}
\end{equation}\label{eq:mbb}

\noindent between $100$--$850\,\mu$m, where $\beta$ is the dust emissivity index. We account for bands in which the galaxy is not detected by extending the formalism of \citet[][]{sawicki12} to the far-IR and sub-mm fluxes. A full description of this methodology can be found in \citet{sawicki12}, but briefly, we measure the goodness-of-fit of greybody templates to the photometry (comprising detections in bands $i$ and non-detections in bands $j$, respectively) by adopting the modified $\chi^2$ statistic:

\begin{equation}
\begin{split}
\chi^2 & = \sum_{i}\biggl(\frac{f_{d,i}-sf_{m,i}}{\sigma_{i}}\biggr)^2\\
       & -2\sum_{j}\ln\int_{-\infty}^{f_{{\rm lim},j}}\exp\biggl[-\frac{1}{2}\biggl(\frac{f-sf_{m,j}}{\sigma_{j}}\biggr)^2\biggr]df
\end{split}
\end{equation}\label{modifiedchi2}

\noindent where $f_{m,i}$ is the model flux density in the $i^{\rm th}$ band, $f_{d,i}$ is the measured flux density in the same band, $\sigma_{i}$ is the uncertainty on the observed flux density, $s$ is the flux scaling between the model and the data (a parameter which is fit numerically\footnote{In the case where the stack is detected in each band, scaling the best-fit template to the photometry is trivial; however in cases with non-detections, we need to numerically find the value of $s$ for which $\partial\chi^2/\partial s=0$}), and $f_{{\rm lim},j}$ is the $1\sigma$ flux uncertainty in the $j^{\rm th}$ band. In the case where the stack is detected in every band (i.e.\,there are no upper limits), the second sum is set to zero, and the fitting reduces to a simple $\chi^2$ determination. 

The functional form of the modified blackbody curve is such that its partial derivatives with respect to $\beta$ and $T_{\rm dust}$ are correlated, and thus exact values for either parameter cannot be analytically determined without first fixing (i.e.\,assuming) the other. While methods exist in the literature to numerically disentangle this correlation \citep[e.g.\,][]{kirkpatrick13, hunt15}, data of high signal-to-noise (${\rm S/N}\geq 10$) are typically required. In the present case, the S/N of our far-IR/sub-mm stacks is insufficient to break this degeneracy; to facilitate comparison with the literature, we therefore fix $\beta = 1.5$, but note that allowing for a range of $\beta=1$--2.5 \citep[as reported in recent works, e.g.\,][]{chapin11,casey11}, would introduce an additional uncertainty on all quantities derived from the SED fits \citep[e.g.\,][]{ibar15}.

We estimate the uncertainties in our isothermal greybody fits using a Markov Chain Monte Carlo (MCMC) approach; we model the stacked photometry for each HiZELS sub-sample using the isothermal grey-body described in Equation\,\ref{eq:greybody} and use the affine-invariant, Markov Chain Monte Carlo sampler, {\sc emcee} \citep{foremanmackey13}. For each set of photometry, we employ 50 ``walkers'' for a combined total of $10^6$ steps. The ``burn-in'' phase is considered over after $5000$ steps of each walker. This is a conservative approach and an investigation of the resultant data indicates that the ``burn-in'' phase is indeed complete and that the chain is well-mixed. The best-fit parameters are taken as those corresponding to the maximum likelihood sample of the chain, and the uncertainties are the $16$--$84^{th}$ percentile of each parameter distribution.

In addition to fitting this isothermal greybody component (from which $T_{\rm dust}$, $M_{\rm dust}$ and $L_{\rm IR}$ are measured), we also fit to the data (including the radio) models from a suite of 184 template SEDs from \citet{chary01}, \citet{draineli07} and \citet{rieke09}, plus the SEDs of the starburst galaxies M\,82, Arp\,220 and SMM\,J21352--0102 \citep{swinbank10a}. In each case, the template SEDs are set to the redshift of the stack, and each template is re-scaled in flux to fit the stacked photometry. The range of best-fitting templates for each stack, which satisfy $\mid \chi^2_{\rm reduced} - \chi^2_{\rm reduced, min}\mid \leq 1$, are shown in Fig.\,\ref{fig:seds}. While the templates implicitly account for both cold and warm dust -- and hence provide the most secure estimate of the total $8$--$1000\,\mu$m luminosity, $L_{8-1000\,\mu{\rm m}}$ -- in order to facilitate comparison with the literature, we hereafter focus primarily on the cold dust component, and take $L_{\rm IR}$ to be the integral under the isothermal greybody curves (unless noted otherwise). The average ratio between the two is $\log_{10}(L_{8-1000\,\mu{\rm m}}) = 1.03\times\log_{10}(L_{\rm IR}$).

We measure dust masses, $M_{\rm dust}$, from our stacked far-IR/sub-mm SEDs via the monochromatic flux density $S_{\nu}$ (at frequency $\nu$). $M_{\rm dust}$ and $S_{\nu}$ are related through the relationship

\begin{center}
  \begin{equation}\label{eq:mdust}
    S_{\nu} \propto \kappa_{\nu}B_{\nu}(T)(1+z)M_{\rm dust}/D_{\rm L}^{2}
  \end{equation}
\end{center}

\noindent where $\kappa_{\nu}$ is the frequency-dependent dust absorption coefficient, $B_{\nu}(T)$ is the Planck function at temperature $T$, and $D_{\rm L}$ is the luminosity distance to the source \citep[e.g.\,][]{casey12a}. In the interest of consistency, we choose to measure $M_{\rm dust}$ from the flux density of the best-fitting modified blackbody at the same rest-frame wavelength ($850$\,$\mu$m) in each case. We measure $S_{850\,\mu {\rm m}}$ in each case by interpolating the best-fit greybody (i.e.\,with $\beta = 1.5$), and fix $\kappa_{850\,\mu{\rm m}}=0.07\pm0.02$\,m$^2$\,kg$^{-1}$, which \citet{james02} find to provide a suitable fit to both dwarf and massive star-forming galaxies in the SCUBA Local Universe Galaxy Survey. We report the implied dust masses in Table\,\ref{tab:hizels-properties}\footnote{Recently, \citet{clark16} measured the equivalent quantity at $500\,\mu$m in 22 massive, dusty galaxies from the \herschel\ Reference Survey, finding $\kappa_{500}=0.051$\,m$^2$\,kg$^{-1}$ (equivalent to $\kappa_{850}\sim 0.02$). Adopting this value instead would increase our dust masses by a factor $\sim2$--$3\times$.}; uncertainties on the dust masses are calculated by propagating the uncertainties on $T_{\rm d}$ from the SED fits through Equation\,\ref{eq:mdust}.

\subsection{The stellar masses of stacked samples of H$\alpha$-emitters}\label{sect:mstar}

In order to relate the far-IR properties of our stacked H$\alpha$-selected samples to their stellar content, it is necessary to determine the ``representative'' stellar mass of each stacked sample. \citet{sobral14} measured the stellar masses of individual HiZELS galaxies via multi-band UV--mid-IR SED fits. While one approach to calculating the ``representative'' stellar mass for our stacks would be to simply calculate the median mass of the galaxies in each sub-sample, this approach does not take in to account the fact that in our SFR-selected sub-samples, the contribution each galaxy makes to the far-IR flux densities is a function of SFR. A more appropriate approach, therefore, is to weight the stellar masses of our galaxies by $L_{{\rm H}\alpha}$, such that the characteristic stellar mass in each stacked sub-sample, $\langle M_\star^{{\rm H}\alpha}\rangle$, is:

\begin{center}
\begin{equation}
\langle M_\star^{{\rm H}\alpha}\rangle = \frac{\sum_{i}L_{{\rm H}\alpha\,,i}M_\star^i}{\sum_{i}L_{{\rm H}\alpha\,,i}}
\end{equation}
\end{center}

These H$\alpha$ luminosity-weighted stellar masses are $\sim 1$--$3\times 10^{10}$\,M$_\odot$, and are around $\sim 7\times$ higher than the simple medians of the stellar masses of the galaxies in each stack. We estimate the typical uncertainties on these weighted masses by bootstrap re-sampling the errors on the numerator and the denominator, and propagating them.

We test the reliability of this weighting scheme by splitting each $>0.2L^\star(z)$ sample in to two smaller sub-samples, comprising the brightest and faintest 50\% of H$\alpha$-luminosity galaxies above $0.2L^\star(z)$ at each redshift. We then stack the far-IR photometry for these and derive $L_{\rm IR}$ for each sub-sample from the stacked photometry as before. We estimate the IR-weighted representative stellar masses of our $>0.2L^\star(z)$ sub-samples from these fits as

\begin{center}
\begin{equation}
\langle M_{\star}^{\rm IR}\rangle = \frac{\langle M_{\star}^{{\rm H}\alpha\,{\rm bright}}\rangle\langle L_{\rm IR}^{{\rm H}\alpha\,{\rm bright}}\rangle + \langle M_{\star}^{{\rm H}\alpha\,{\rm faint}}\rangle\langle L_{\rm IR}^{{\rm H}\alpha\,{\rm faint}}\rangle}{\langle L_{\rm IR}^{{\rm H}\alpha\,{\rm bright}}\rangle + \langle L_{\rm IR}^{{\rm H}\alpha\,{\rm faint}}\rangle}
\end{equation}
\end{center}

We measure $\langle M_\star^{\rm IR}\rangle =(7.5\pm 1.9)\times 10^{9}$\,M$_\odot$,$(7.3\pm 2.0)\times 10^{9}$\,M$_\odot$,$(10.3\pm 1.7)\times 10^{9}$\,M$_\odot$ and $(11.6\pm 4.8)\times 10^{9}$\,M$_\odot$ in each of our four samples, with ascending redshift, in excellent agreement with the H$\alpha$-weighted stellar masses at $z=0.84$ and $z=1.47$ (within 25\%), and within a factor $\sim 2\times$ of the H$\alpha$-weighted mass at $z=2.23$. At $z=0.40$, the discrepancy between the H$\alpha$-weighted and IR-weighted stellar masses is a factor $\sim4\times$; however much of this may be explained by the small number of $>0.2L^\star_{{\rm H}\alpha}(z)$ HiZELS galaxies at this redshift (52), which results in the stacked SEDs for the two halves of this sample being characterised by upper-limits at 100, 160 and 350\,$\mu$m. This hence leads to large uncertainties in $L_{\rm IR}$. We hereafter use our H$\alpha$-weighted stellar masses to characterise our samples, and report these in Table\,\ref{tab:stack-properties}.

\section{Results and discussion}\label{sec:results}
\subsection{SED fits}\label{sec:sedfits}
\begin{figure*}
\centerline{\psfig{file=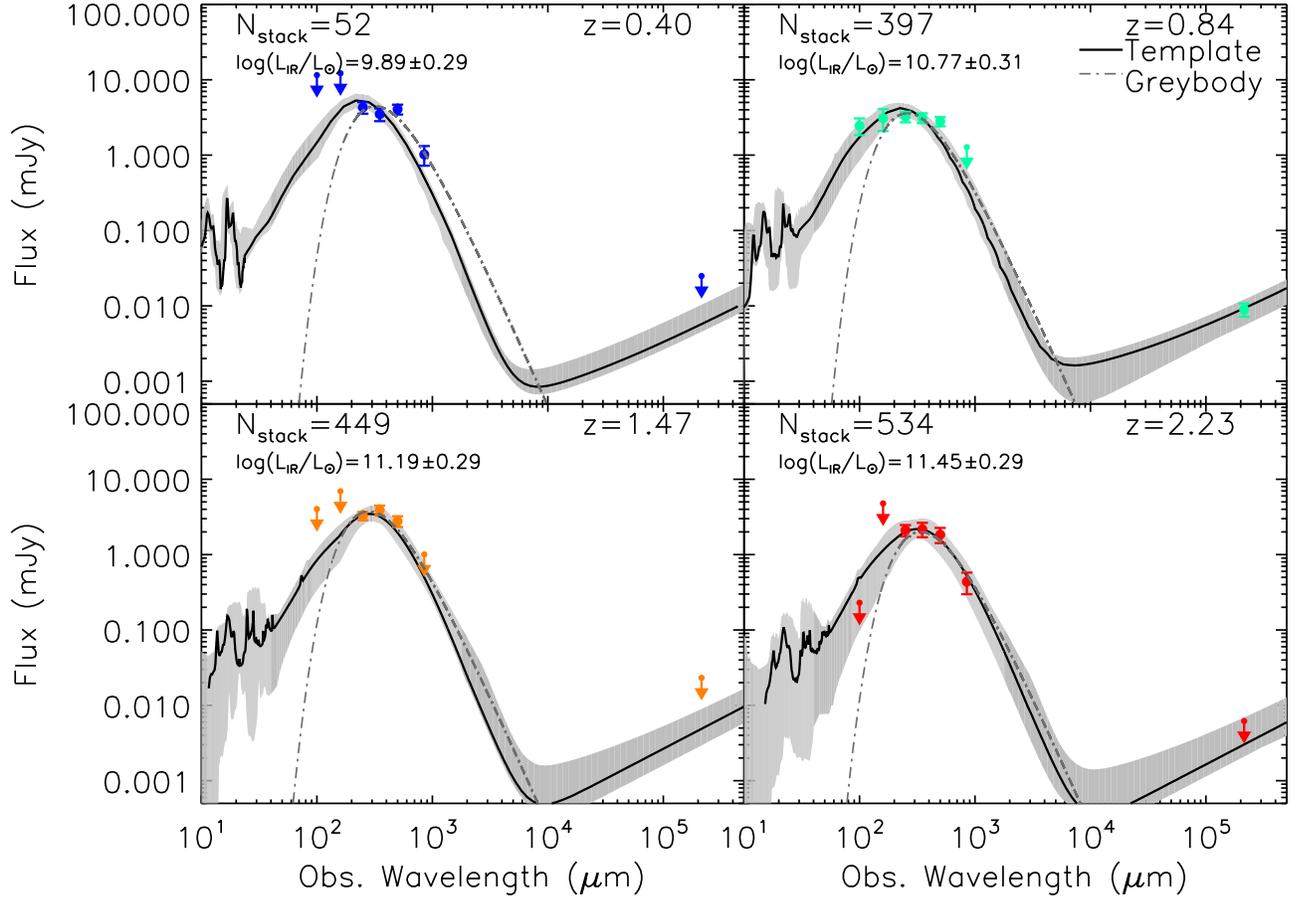,height=\textwidth,angle=90}}
\caption[Short captions]{Far-infrared/sub-mm spectral energy distributions of stacks comprising all HiZELS star-forming galaxies with $L_{{\rm H}\alpha} \geq 0.2L^{\star}_{{\rm H}\alpha}(z)$, in each of the four redshift slices. Stacked photometry are shown with coloured points; non-detections are represented as $3\sigma$\ upper limits with arrows. We show the best-fitting template SED for each stack, along with the range of templates within $\chi^2_{\rm red}=1$ as a grey band. The best-fitting isothermal grey-body is shown with a dot-dashed line (see \S\,\ref{sec:sedfitting} for details). The templates implicitly account for the the effects of warm dust (which dominates the SED at $\lambda_{\rm rest}\lesssim 100\,\mu$m wavelengths); however in order to facilitate comparison with the literature, we take $L_{\rm IR}$ to be the the integral under the isothermal greybody, representing the dominant cold dust component only.}
\label{fig:seds}
\end{figure*}

\begin{table*}
\centering
\caption[Stacked photometry]{Stacked photometry}
\label{tab:hizels-photometry}
\begin{tabular}{lccccccccc}
\hline
\multicolumn{1}{l}{} &
\multicolumn{1}{c}{$S_{100\,\mu{\rm m}}$} &
\multicolumn{1}{c}{$S_{160\,\mu{\rm m}}$} &
\multicolumn{1}{c}{$S_{250\,\mu{\rm m}}$} &
\multicolumn{1}{c}{$S_{350\,\mu{\rm m}}$} &
\multicolumn{1}{c}{$S_{500\,\mu{\rm m}}$} & 
\multicolumn{1}{c}{$S_{850\,\mu{\rm m}}$} & 
\multicolumn{1}{c}{$S_{1.4\,{\rm GHz}}$} & \\
\multicolumn{1}{l}{} &
\multicolumn{1}{c}{$({\rm mJy})$} &
\multicolumn{1}{c}{$({\rm mJy})$} &
\multicolumn{1}{c}{$({\rm mJy})$} &
\multicolumn{1}{c}{$({\rm mJy})$} & 
\multicolumn{1}{c}{$({\rm mJy})$} & 
\multicolumn{1}{c}{$({\rm mJy})$} & 
\multicolumn{1}{c}{$(\mu{\rm Jy})$} &\\
 \hline
$z=0.40$&$<4.4$&$<7.6$&$4.3\pm 0.7$&$3.5\pm 0.7$&$4.1\pm 0.6$&$1.0\pm 0.3$&$<13.3$\\
$z=0.84$&$2.5\pm 0.6$&$3.1\pm 1.0$&$3.2\pm 0.5$&$3.1\pm 0.5$&$2.8\pm 0.4$&$<0.4$&$8.9\pm 1.8$\\
$z=1.47$&$<1.7$&$<2.7$&$3.3\pm 0.4$&$3.9\pm 0.6$&$2.8\pm 0.5$&$<0.5$&$<7.9$\\
$z=2.23$&$<1.4$&$<2.4$&$2.1\pm 0.4$&$2.2\pm 0.5$&$1.8\pm 0.4$&$0.4\pm 0.1$&$<6.1$\\
\hline
\end{tabular}
\\ {\small Notes: Non-detections are represented as $3\sigma$\ limits}
\end{table*}

We show our stacked photometry in Fig.\,\ref{fig:seds}, along with the best-fitting template and greybody SEDs. Each of our H$\alpha$-selected sub-samples is detected in all three SPIRE bands, while the PACS data (which cover only the COSMOS field) typically present only upper limits (with $>3\sigma$ detections in both bands only at $z=0.84$). Properties measured from these SED fits are presented in Table\,\ref{tab:hizels-properties}. The measured infrared luminosities of our H$\alpha$-selected galaxies increase from $\log_{10}[L_{\rm IR}/{\rm L}_{\odot}]=9.9$--$11.5$ with redshift, while the grey-body dust temperatures increase from $T_{\rm dust}=14$--$34$\,{\sc k}. 

In our three higher redshift bins, we see that the best-fitting isothermal greybody is consistent with the $\pm1\sigma$ range of template SED fits. At $z=0.40$, the isothermal greybody is $\sim8$\,{\sc k} colder than the best-fitting template ($T_{\rm greybody}=14\pm3$\,{\sc k}, versus $T_{\rm template}=22\pm1$\,{\sc k}). The coldest available template in the template library is $20$\,{\sc k}, which indicates that much of this discrepancy can be accounted for by the template library's inadequate sampling of sufficiently cold dust temperatures. The $z=0.40$ greybody fit is also significantly colder than any of the 500 individually far-IR detected $z=0$--$3$ star-forming galaxies observed by \citet{genzel15}. However our $z=0.40$ H$\alpha$-selected galaxies are typically a factor $10\times$\ less massive than the galaxies detected by \citet{genzel15}, and have a lower ${\rm SFR}\sim 2$\,M$_{\odot}$\,yr$^{-1}$.

We see no evidence for strong evolution of $M_{\rm dust}$ with redshift, with each of our stacked SEDs for H$\alpha$-selected star-forming galaxies being consistent with the mean dust mass of $(1.6\pm 0.1)\times10^{8}$\,M$_\odot$ measured by \citet{rowlands14} for a sample of $z<0.5$ dusty star-forming galaxies. 

\begin{table*}
\centering
\caption[Properties of stacked H$\alpha$-selected samples]{Properties of stacked H$\alpha$-selected samples}
\label{tab:hizels-properties}
\begin{tabular}{lccccccccc}
\hline
\multicolumn{1}{l}{} &
\multicolumn{1}{c}{$\log_{10}(L_{\rm IR})$} & 
\multicolumn{1}{c}{$T_{\rm dust, BB}$} &
\multicolumn{1}{c}{$T_{\rm dust, Tem}$} &
\multicolumn{1}{c}{$\log_{10}(L_{\rm IR}/L_{{\rm H}\alpha})$} & 
\multicolumn{1}{c}{$M_{\rm dust}$} & 
\multicolumn{1}{c}{$A_{{\rm H}\alpha, {\sc cont}}$} & 
\multicolumn{1}{c}{$A_{{\rm H}\alpha,{\rm ext}}$} & \\
\multicolumn{1}{l}{} &
\multicolumn{1}{c}{(L$_\odot$)} &
\multicolumn{1}{c}{({\sc k})} &
\multicolumn{1}{c}{({\sc k})} &
\multicolumn{1}{c}{} & 
\multicolumn{1}{c}{$(\times10^8\,{\rm M}_{\odot})$} & 
\multicolumn{1}{c}{} &
\multicolumn{1}{c}{} &\\
\hline
$z=0.40$  &$9.89\pm 0.29$&$14\pm 3$&$22\pm 1$&$-1.80\pm 0.05$&$4.5\pm 2.4$&$0.92$&$0.70$& \\
$z=0.84$  &$10.77\pm 0.31$&$25\pm 6$&$32\pm 6$&$-2.16\pm 0.06$&$1.3\pm 0.8$&$1.18$&$0.85$& \\
$z=1.47$  &$11.19\pm 0.29$&$27\pm 7$&$30\pm 7$&$-2.01\pm 0.05$&$2.2\pm 1.1$&$1.05$&$0.78$& \\
$z=2.23$  &$11.45\pm 0.29$&$34\pm 8$&$34\pm 7$&$-2.15\pm 0.06$&$1.2\pm 0.6$&$0.92$&$0.70$& \\
\hline
\end{tabular}
\\ {\small Notes: Uncertainties on all far-IR derived properties are measured by adding (in quadrature) the statistical uncertainties from the isothermal greybody fits to the model-dependent systematic, measured as the offset between the isothermal and dual-temperature fits (\S\,\ref{sec:sedfitting}).}
\end{table*}

\subsection{Extinction properties of H$\alpha$ emitters}\label{sec:extinction}

In applying individual dust corrections to each of our H$\alpha$-selected star-forming galaxies based on the Calzetti reddening law, we have potentially altered the compositions of the samples in our $>0.2L^\star(z)$ stacks relative to samples selected with no extinction correction applied; any non-zero dust correction will tend to scatter galaxies with lower observed H$\alpha$ fluxes above our luminosity cut-offs which would otherwise not have surpassed this threshold.

To investigate whether this effect has in any way biased our conclusions, we now perform a series of tests comparing the properties of our dust-corrected sub-samples with analogous sub-samples to which \textit{no} dust correction is applied. The median $E(B-V)$ of galaxies derived via multi-wavelength SED fits in the \textit{corrected} sub-samples are $0.35$ at $z=0.40$, $0.45$ at $z=0.84$, $0.40$ at $z=1.47$ and $0.35$ at $z=2.23$, corresponding to $A_V=1.12$, $1.44$, $1.28$ and $1.12$, respectively. The H$\alpha$ extinctions ($A_{{\rm H}\alpha}$) measured from this analysis (via the process outlined in \S\,\ref{sec:dust-corrections}), along with the number of HiZELS galaxies above our $0.2L^\star(z)$ thresholds are listed in Table\,\ref{tab:hizels-properties}. 

We also select (and stack) comparison sub-samples consisting of the same number of galaxies at each redshift, representing those galaxies with the highest \textit{uncorrected} $L_{{\rm H}\alpha}$, and fit far-IR SEDs to the photometry in the manner set out in \S\,\ref{sec:sedfits}. 

\begin{figure*}
\centerline{\psfig{file=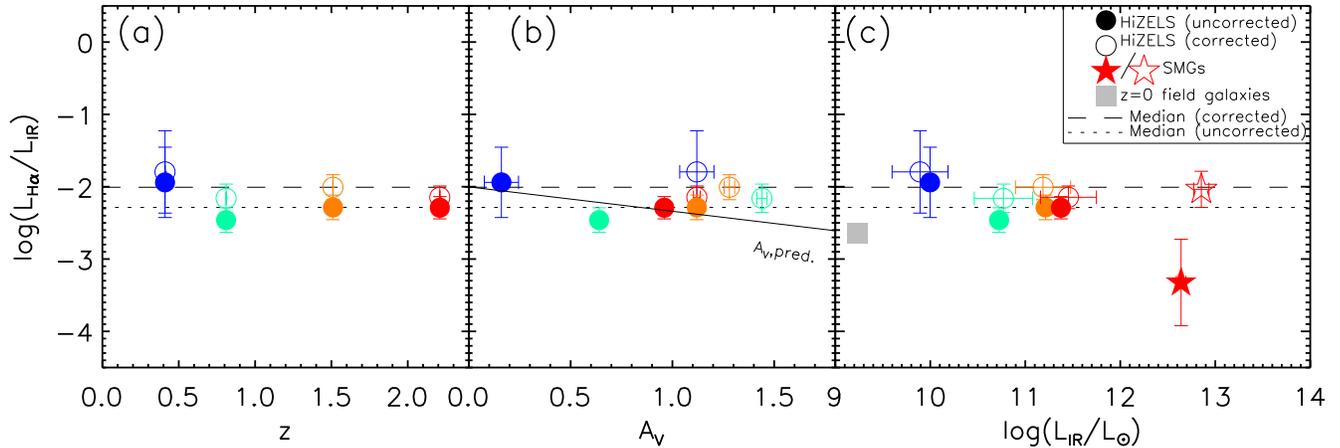,width=\textwidth}}
\caption[AVplot]{$(a)$ We show the relationship between the extinction proxy $\log_{10}[L_{{\rm H}\alpha}/L_{\rm IR}]$ and redshift. We see no evolution in the ratio $\log_{10}[L_{{\rm H}\alpha}/L_{\rm IR}]$ -- using samples selected via either their extinction corrected (see \S\,\ref{sec:extinction}) or uncorrected $L_{{\rm H}\alpha}$ -- with redshift, suggesting that our selection of galaxies above an evolving H$\alpha$ luminosity cut yields sub-samples with similar ISM properties at different redshifts. $(b)$ We compare $\log_{10}[L_{{\rm H}\alpha}/L_{\rm IR}]$ as a reddening proxy with the measured stellar extinctions, $A_V$, finding our extinction-corrected H$\alpha$-selected stacks to have similar $A_V$. We plot the empirically-derived relationship between the two quantities, based on the $L_{{\rm H}\alpha}$- and $L_{\rm IR}$-to-SFR conversion factors of \citet{kennicutt12} and \citet{kennicutt98}, plus the \citet{calzetti00} reddening law, and find that this exactly intersects our sample. $(c)$ If $\log_{10}[L_{{\rm H}\alpha}/L_{\rm IR}]$ is a suitable proxy for the dust extinction, then we may expect to see a negative correlation between this ratio and $L_{\rm IR}$, given the latter's connection with the cool dust component of the ISM. We see no strong trend, which may indicate that very heavily extinguished (low $L_{{\rm H}\alpha}/L_{\rm IR}$), dusty star-forming galaxies do not dominate our stacks. We show the median results from the dust-corrected samples of local field galaxies \citep{kewley02} and SMGs \citep[][open star]{takata06}, as well as the non-corrected SMG sample of \citet[][filled star]{swinbank04}. In each panel, the median corrected and uncorrected $\log_{10}[L_{{\rm H}\alpha}/L_{\rm IR}]$ of our H$\alpha$-selected sub-samples are also shown. The colour-coding is the same as in Fig.~\ref{fig:seds}.}
\label{fig:avplots}
\end{figure*}

In Fig.~\ref{fig:avplots} we plot the ratio of the median H$\alpha$ luminosity of galaxies in each of these sub-samples to $L_{\rm IR}$, as measured from the best-fit isothermal greybodies. If no $A_{{\rm H}\alpha}$ correction is applied, the ratio $\log_{10}[L_{{\rm H}\alpha}/L_{\rm IR}]$ provides a proxy for the amount of reddening due to dust. We see that this ratio remains approximately constant from $z=0.40$--$2.23$ for both the corrected and uncorrected sub-samples. This argues for no significant evolution in the reddening of ``typical'' H$\alpha$-selected galaxies as a function of redshift \citep[see also][]{garn10,sobral12,ibar13}, with a scalar offset between the corrected and uncorrected sub-samples of $\Delta\log_{10}[L_{{\rm H}\alpha}/L_{\rm IR}]\sim 0.2$.

Next, we compare $\log_{10}[L_{{\rm H}\alpha}/L_{\rm IR}]$ with the observed median $A_V$ of each sub-sample. The lack of any strong correlation between the ratio $\log_{10}[L_{{\rm H}\alpha}/L_{\rm IR}]$ and $A_V$ seen in Fig.~\ref{fig:avplots} for our uncorrected sub-samples is consistent with the lack of correlation seen in corrected sub-samples, and again suggests that stacking on sub-populations defined relative to $L^{\star}_{{\rm H}\alpha}$ selects similar galaxies at all redshifts (albeit, with a tendency to select intrinsically more luminous galaxies at higher redshift as $L^{\star}(z)$ evolves). However, the sub-samples selected on the basis of their \textit{uncorrected} H$\alpha$ luminosities have a much lower typical $A_V$ than those in the corrected sub-samples. This suggests that applying individual dust-corrections to the H$\alpha$ luminosities (measured from their stellar continuum fits) prior to stacking is an important step, as it allows high-SFR (but dusty) galaxies to satisfy our $0.2L_{{\rm H}\alpha}(z)$ criterion, which otherwise would be excluded on the basis of their low (observed) H$\alpha$ fluxes.

In Fig.~\,\ref{fig:avplots}, as a sanity check, we also show the implied relationship between $\log_{10}[L_{{\rm H}\alpha}/L_{\rm IR}]$ and $A_{V\,{\rm pred}}$, where $A_{V\,{\rm pred}}$ is the predicted extinction derived (for the uncorrected sample) by equating the $L_{\rm IR}$-to-SFR and $L_{{\rm H}\alpha}$-to-SFR indicators of \citet{kennicutt12} and \citet{kennicutt98}\footnote{the latter corrected by a factor $1.6\times$ to account for the change from a \citet{salpeter55} to a \citet{chabrier03} IMF.}, respectively, and solving for the ``intrinsic'' luminosity ratio that would be seen, if there were no dust extinction:

\begin{center}
\begin{eqnarray}\label{eq:kennicutt}
\log({\rm SFR}/{\rm M}_\odot \,{\rm yr}^{-1})=\log(\lhamath/{\rm ergs\,s}^{-1}) - 41.36
\end{eqnarray}
\vspace*{-6mm}
\begin{eqnarray*}
 = \log(L_{\rm IR}/{\rm ergs\,s}^{-1}) - 43.41
\end{eqnarray*}
\end{center}

\vspace*{4mm}
\noindent such that

\vspace*{4mm}
\begin{center}
\begin{equation}\label{eq:kennicutt2}
\log(\lhamath/L_{\rm IR})\big|_{\rm int} = -2.05
\end{equation}
\end{center}

Next, we compute $A_{{\rm H}\alpha}$ (again, for the uncorrected sub-samples) by comparing the observed and intrinsic \lha-to-$L_{\rm IR}$ ratios

\begin{center}
\begin{equation}\label{eq:luminosity-ratio-comparison}
A_{{\rm H}\alpha} \equiv \frac{\log(\lhamath/L_{\rm IR})\big|_{\rm obs}-\log(\lhamath/L_{\rm IR})\big|_{\rm int}}{0.4}
\end{equation}
\end{center}

\noindent before finally deriving $A_{V\,{\rm , pred}}$ via the \citet{calzetti00} law, $A_{V\,{\rm , pred}} = A_{{\rm H}\alpha}R'_V/\kappa(\lambda)$. The predicted trend is for low-extinction galaxies to have higher H$\alpha$-to-far-IR ratios, and conversely for higher-extinction galaxies ($A_V\gtrsim 1$) to have lower H$\alpha$-to-far-IR ratios, due to the increased absorption of H$\alpha$ photons in the dustier regions which give rise to far-IR emission. 

We see from Fig.~\ref{fig:avplots} that our four (extinction-corrected) stacked HiZELS samples occupy a region of moderate $A_V\sim 1$--$1.5$. The un-corrected H$\alpha$-selected stacks, as previously noted, have lower $A_V$ than the corrected stacks. The likely explanation for this is simply that only those galaxies with low $A_V$ are likely to exceed $0.2L_{{\rm H}\alpha}^\star(z)$ if no correction is applied, and that it is the inclusion of intrinsically luminous (but heavily extinguished) galaxies -- after applying a dust correction -- that raises the median $A_V$ in the corrected sub-samples. We also note that the difference in $A_V$ between the corrected/uncorrected sub-samples at $z=0.40$ is greater than the difference between any other two sub-samples at a given redshift. This is most likely because at $z=0.40$, the volume probed by our H$\alpha$ observations is smaller than at any other redshift, meaning fewer galaxies exceed $0.2L_{{\rm H}\alpha}^\star$ than at any other redshift -- hence the up-scattering of a small number of additional, heavily extinguished galaxies above $0.2L_{{\rm H}\alpha}^\star$ by performing a full dust correction produces larger random shifts in the derived properties of the $z=0.40$ sub-sample than at other redshifts, where the larger survey volumes mitigate the effect of up-scattering a small number of heavily extinguished galaxies.

In Fig.~\ref{fig:avplots}, we also show $\log_{10}[L_{{\rm H}\alpha}/L_{\rm IR}]$ as a function of $L_{\rm IR}$, along with comparison samples from the literature. We see no evidence of strong trends in $L_{{\rm H}\alpha}/L_{\rm IR}$ as a function of $L_{\rm IR}$, in either the corrected or uncorrected HiZELS sub-samples, with $\log_{10}[L_{{\rm H}\alpha}/L_{\rm IR}]$ lying within $\pm 1\sigma$ of the median at each redshift, regardless of whether we apply individual $A_{{\rm H}\alpha}$ corrections or not. Hence, although we cannot claim to observe \textit{strong} evolution in $L_{{\rm H}\alpha}/L_{\rm IR}$ with $L_{\rm IR}$ -- which we would expect, if higher-$L_{\rm IR}$ galaxies were found to be dustier on average -- the relationship between $L_{{\rm H}\alpha}/L_{\rm IR}$ and $L_{\rm IR}$ is at least \textit{weakly} consistent with our expectations. These results suggest that while our extinction-corrected H$\alpha$-selected stacks do contain some IR-luminous, dusty galaxies, they are not dominated by extreme sources.

\subsection{Luminosity-temperature relation}\label{sec:lirtd}

\begin{figure*}
\includegraphics[height=\textwidth,angle=90]{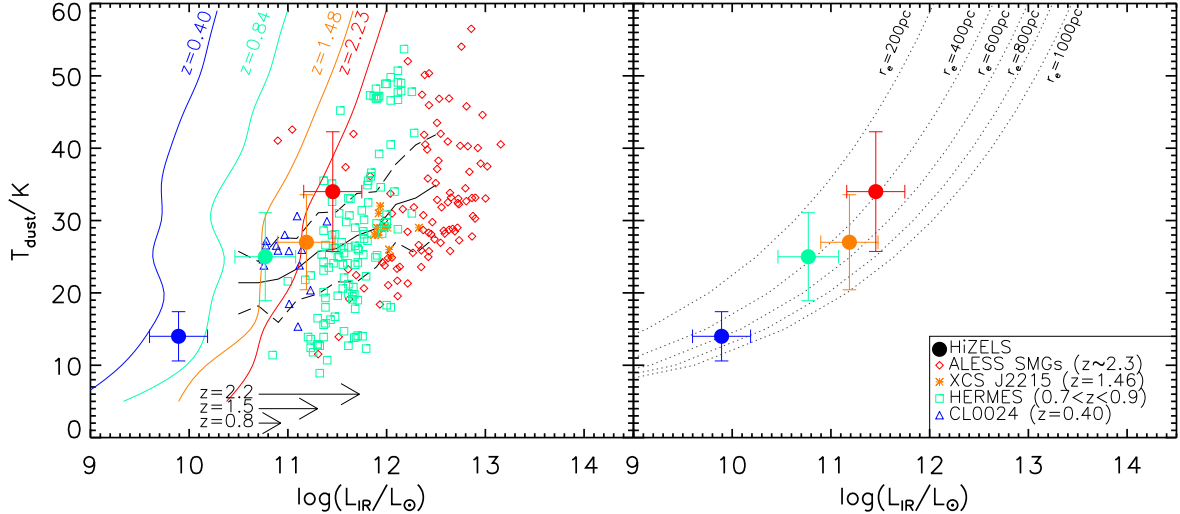}
\caption[Short captions]{\textit{Left:} Dust temperature, $T_{\rm dust}$ versus the modified blackbody luminosity, $L_{\rm IR}$, for the four extinction-corrected, H$\alpha$-selected HiZELS samples, using properties measured from the best-fitting SEDs (\S\,\ref{sec:stacking}). HiZELS stacks are plotted with large filled symbols, and use the same colour coding as in Fig.~\ref{fig:seds}. We show the relation for $0.1<z<2$ IR-luminous star-forming galaxies (black line with $\pm2\sigma$ scatter) from \citet{symeonidis13}. Typical selection boundaries are shown with colour-coded lines for each redshift (see text for details). The arrows at the bottom of the plot indicate the expected increase in $L^{\star}_{\rm IR}(z)$ relative to $z=0.40$, given the evolution in $L_{{\rm H}\alpha}^\star(z)$, and the common scaling between \lha\ and L$_{\rm IR}$ proposed in \citet{kennicutt12}. This evolution roughly matches that seen in our samples, in the sense that straightforwardly converting $L_{{\rm H}\alpha}^\star(z)$ to ${\rm SFR}^\star(z)$ to $L_{\rm IR}^\star(z)$ using the scaling relations of \citet{kennicutt12} places the arrow heads at infrared luminosities that are close to those we measure from the stacks.  We also show comparison samples observed in the far-IR at the redshifts of our HiZELS stacks, including a population of ALMA-detected $z>2$ submillimetre galaxies \citep{swinbank14}, submillimetre-detected starburst galaxies in the $z=1.46$ cluster, XCS\,J2215 \citep{ma15}, a population of $\langle z \rangle \sim 0.8$ far-infrared selected starbursts from \citet{casey12a}, and starburst galaxies in the core of the $z=0.4$ cluster CL0024+17 \citep{johnson16}. \textit{Right:} As the left hand plot, but with the comparison samples and the local relation removed. If the dust is well described by a greybody SED, then it ought to obey a modified Stefan-Boltzmann (SB) law, $L_{\rm IR}\propto R^2 T_{\rm dust}^4$. We show curves of constant size, using the prescription outlined in \S\,\ref{sec:sizes}, finding that the dust emission from our H$\alpha$ stacks is well-described as having a constant characteristic size $\sim500$\,pc, with increasing temperature and luminosity at higher redshifts. We note that this size is $\sim 10\times$ smaller than the median stellar sizes of HiZELS galaxies, which may suggest the dust is bound in clumpier knots and filaments, and does not directly trace the stellar emission.}
\label{fig:lir-v-td}
\end{figure*}

In Fig.~\ref{fig:lir-v-td}, we show the measured infrared luminosities and dust temperatures for each of the four HiZELS stacks. We see that both $L_{\rm IR}$ and $T_{\rm dust}$ appear to increase with redshift in our sample from $z=0.40$ to $2.23$. Due to the evolving H$\alpha$ luminosity threshold -- and the fact that both \lha\ and $L_{\rm IR}$ correlate with the SFR -- we expect $L_{\rm IR}$ of our stacked H$\alpha$-selected galaxies to increase with redshift,. This behaviour is indeed seen, with $L_{\rm IR}$ increasing from $10^{9.9\pm 0.3}$\,L$_\odot$ to $10^{11.5\pm 0.3}$\,L$_{\odot}$ between $z=0.40$ and $2.23$ -- an increase of a factor $\sim40\times$, compared to the increase of a factor $\sim10\times$ in $L^{\star}_{{\rm H}\alpha}(z)$. 

We determine the effective selection boundaries for our HiZELS samples on the plot of $T_{\rm dust}$ against $L_{\rm IR}$ by generating $1000$ isothermal greybody SEDs at each redshift (with fixed $\beta=1.5$), in increments of $T_{\rm dust}=5\,${\sc k} between $5$--$60$\,{\sc k}. At each redshift, and at each temperature, we find the lowest-$L_{\rm IR}$ greybody SED that is above the measured $3\sigma$\ detection limits of at least two of the SPIRE band stacked images. Our HiZELS stacked sub-samples are typically $3\times$ more luminous than the selection limits at the corresponding temperature and redshift.

For comparison, we plot the $L_{\rm IR}$--$T_{\rm dust}$ relation from \citet{symeonidis13}, defined for $0.1<z<2$ LIRG/ULIRG galaxies in the \herschel\ Multi-Tiered Extragalactic Survey \citep[HerMES;][]{oliver10} and PACS Evolutionary Probe \citep[PEP;][]{lutz11} surveys. In addition, we also populate Fig.~\ref{fig:lir-v-td} with data from comparison samples at similar redshifts to the HiZELS slices taken from the literature. These include  the $z=0.40$ galaxy cluster Cl\,0024+16 \citep{johnson16}, $\langle z \rangle = 0.8$ \herschel-SPIRE detected SFGs from \citet{casey12a}, Ultra-Luminous InfraRed Galaxy (ULIRG) starbursts in the $z=1.46$ cluster XCS\,J2215 \citep{ma15}, and also $z=2$--$3$ ALMA-detected SMGs in the Extended \textit{Chandra} Deep Field South \citep[ALESS;][]{swinbank14}. 

We see that our H$\alpha$-selected stacks are typically a little warmer (at fixed $L_{\rm IR}$) than the \citet{symeonidis13} local relation, but similar in temperature to the comparison samples (which comprise individually far-IR detected galaxies) at each redshift. The ability of our stacking approach to extend the $L_{\rm IR}/T_{\rm dust}$ relationship to lower luminosities compared with the individual detections in the comparison samples is also apparent; at $z=0.40$ and $0.84$, our H$\alpha$-selected stacks are both lower-luminosity and (a little) cooler than the individually-detected galaxies in their respective comparison samples. At $z=1.47$ and $2.23$ our stacked H$\alpha$-selected samples are roughly an order of magnitude fainter than the individually-detected galaxies in their respective comparison samples, but have similar dust temperatures.

\subsection{Dust sizes}\label{sec:sizes}
To interpret the locations of the HiZELS stacks on the $L_{\rm IR}$--$T_{\rm dust}$ plot, we employ a simple model that relates the luminosity, temperature, and expected size of the far-IR emitting regions in our sample. For a spherical blackbody source, the Stefan-Boltzmann law ($L = 4\pi R^2\sigma T^4$, where $\sigma$ is the Stefan-Boltzmann constant and $R$ is the radius of the emitting source) provides the natural framework within which to interpret the $L_{\rm IR}$--$T_{\rm dust}$ relation. In the case of greybody emission, a form of the Stefan-Boltzmann law will apply, in which the Stefan-Boltzmann constant, $\sigma$, is replaced with $\tilde{\sigma}(T,\beta,\lambda_0)$ for a given set of dust properties. Choosing $\beta = 1.5$ (as we did during the SED fitting), and fixing $\lambda_0 \sim 100\,\mu$m as the reference wavelength at which the dust opacity is unity, \citet{yan16} investigated this relationship in a sample of high-redshift, dusty star-forming ULIRGs, finding that for dust temperatures between $10$--$100$\,{\sc k}, $\tilde{\sigma}$ can be approximated as: 

\begin{center}
\begin{equation}
\frac{\tilde{\sigma}(T_{\rm d})}{\sigma} = 10^{-3}\bigl(-3.03T_{\rm d}^{1.5}+45.55T_{\rm d}-127.53\bigr)
\end{equation}
\end{center}

We can use this to estimate the ``effective radii'' of the dust-emitting regions of our H$\alpha$-selected stacks via their locations on the $L_{\rm IR}$--$T_{\rm dust}$ plane as:

\begin{center}
\begin{equation}\label{eq:boltzmann}
R_{\rm eff} = \sqrt{\frac{L_{\rm IR}}{4\pi\tilde{\sigma}T_{\rm d}^4}}
\end{equation}
\end{center}

The effective radii for the dust-emitting regions of the H$\alpha$-selected galaxies in our stacks are between $0.4$--$0.6$\,kpc. We can compare these to the stellar sizes from \citet{stott13}, who measured the typical sizes of H$\alpha$-selected galaxies by fitting S\'{e}rsic profiles to their rest-frame optical continuum emission. The dust sizes we derive from our analysis are, at all redshifts, nearly an order of magnitude smaller than the stellar continuum sizes, $\langle r_e \rangle = 3.6\pm 0.3$\,kpc. Such a discrepancy may indicate that the bulk of the dust in H$\alpha$-selected star-forming galaxies is not distributed smoothly throughout the ISM, but is instead bound up in (one or more) dense clumps. Direct observational evidence for the existence of dense, star-forming clumps within the ISMs of high-redshift galaxies has traditionally been limited to extreme starbursting systems \citep[e.g.\,][]{swinbank10a, hodge12}, however recent work by \citet{zanella15}, based on spatially-resolved maps of [O{\sc ii}], [O{\sc iii}] and H$\beta$ line emission in a galaxy cluster at $z=1.99$, suggests that similar structures may exist in more modestly star-forming systems as well. We note that a putative dust size of $\sim 500$\,pc corresponds to an angular scale $\sim0.06''$ at $z=2.23$, which is far below the angular resolution limit of \herschel\, and for a galaxy at $\log_{10}[L_{\rm IR}/L_\odot]\sim 11$, would require a significant investment of long-baseline ALMA time to measure directly.

\section{The evolving gas content of the Universe}\label{sec:madau}

In order to understand galaxy evolution, it is necessary to link populations of galaxies at high and low redshift. One means by which high-redshift galaxies can be linked to local populations is by determining the typical timescale on which a galaxy at a given redshift forms stars, and exhausts its gas supply. The length of time a star-forming galaxy can support its present rate of star-formation is determined by its gas depletion timescale ($\tau_{\rm dep}$), defined -- in the absence of inflows/outflows of material -- as the ratio of the available molecular gas supply ($M_{{\rm H}_2}$) to the rate at which that gas is being converted into stars (i.e.\,the SFR).

As noted previously, direct observations of the atomic and molecular gas content of distant galaxies are expensive in terms of telescope time, making them unfeasible for large samples. However, we can obtain indirect constraints on the total gas masses of our H$\alpha$-selected galaxies by invoking the common assumption that the total gas mass ($M_{\rm gas}\equiv M_{\rm H{\sc I}}+M_{{\rm H}_2}$) is proportionally linked to $M_{\rm dust}$ (which we have measured from our far-IR stacks) via a gas-to-dust ratio, $\delta_{\rm GDR}$. 

In the local Universe, star-forming galaxies of approximately solar metallicity, $Z\equiv 12+\log{({\rm O}/{\rm H}}) = 8.7$ \citep{asplund09}, have $\delta_{\rm GDR}\sim 140$ \citep{draine07}. More recent work by \citet{leroy11}, across a more diverse sample of local galaxies, has found that this ratio is sensitive to (and scales inversely with) the metallicity, as 

\begin{center}
\begin{equation}\label{eq:leroy}
\log_{10}\delta_{\rm GDR} = (9.4\pm 1.1) - (0.85\pm 0.13)\bigl[12+\log_{10}({\rm O}/{\rm H})\bigr]
\end{equation}
\end{center}

In order to constrain the gas-to-dust ratio for our H$\alpha$-selected samples, it is therefore necessary to first estimate their metallicities. Metallicities of extragalactic sources are commonly estimated via emission line diagnostics, e.g.\ the [O{\sc iii}]/[N{\sc ii}] ratio \citep{pettini04}, however obtaining such line diagnostics for a statistically large sample is relatively expensive in terms of telescope time. \citet{stott13} carried out multi-object spectrometer observations of the [O{\sc iii}] and [N{\sc ii}] lines for a sub-set of 381 HiZELS galaxies, in order to calibrate the ``fundamental metallicity relation'' (FMR), a 3D surface relating a galaxy's metallicity with its (much more easily observable) stellar mass and SFR.

\citet{schreiber16} perform a similar analysis on a sample of $z=0.7$--$1.3$ star-forming galaxies, selected from the CANDELS field using the FMR calibration of \citet{kewley08}, in which they measure $\delta_{\rm GDR}=150$--$380$. The implied $\log_{10}[M_{\rm gas}/{\rm M}_\odot ]=9$--$10$ for their stellar mass-selected sample galaxies is in good (within $\sim 30\%$) agreement with the gas masses they measured directly via \co\ $+$H{\sc i} spectroscopy. While a direct comparison of the results of the their mass-selected sample with those of our SFR-selected sample is difficult, the broad conclusion from \citet{schreiber16} -- that in spite of the potential uncertainties in inferring $M_{\rm gas}$ from $M_{\rm dust}$ via a metallicity-dependent $\delta_{\rm GDR}$, the derived gas masses correlate well with those obtained from spectroscopy -- is likely to hold for our H$\alpha$-selected sample as well.

\subsection{The metallicities of H$\alpha$-selected galaxies}\label{sec:fmr}

We measure the metallicities of our stacked H$\alpha$-selected samples by exploiting the fundamental metallicity relation (FMR) -- a 3D plane defined by \citet{mannucci10} as

\begin{center}
\begin{eqnarray}\label{eq:fundamentalplane}
12+\log({\rm O/H}) = a_0 + a_1m + a_2s + a_3m^2
\end{eqnarray}
\vspace{-7mm}
\begin{eqnarray*}
+a_4ms + a_5s^2
\end{eqnarray*}
\end{center}

\noindent where $m=\log_{10}[M_{\star}/{\rm M}_{\odot}]-10$, and $s=\log_{10}[{\rm SFR}/{\rm M}_\odot\,{\rm yr}^{-1}]$.  Recently, \citet{stott13}, used Subaru FMOS observations of the [N{\sc ii}]-to-H$\alpha$ line ratio to investigate this metallicity relationship in a sub-sample of the HiZELS galaxies, comprising 381 bright H$\alpha$-selected galaxies at $z=0.84$ and $1.47$. They found that the bright H$\alpha$-selected galaxies occupy an FMR that is relatively flat across the mass range of their sample, and depends primarily on SFR, with best-fit coefficients $a_0=8.77$, $a_1=0.00$, $a_2=-0.055$, $a_3=0.00$, $a_4=0.019$, and $a_5=-0.101$.

We measure $\langle {\rm SFR}_{{\rm H}{\alpha}}\rangle$ for our stacked sub-samples from the median dust-corrected H$\alpha$ luminosities (Table\,\ref{tab:stack-properties}), and use the H$\alpha$-weighted stellar masses, $\langle M_\star\rangle$, of each stacked sub-sample (measured in \S\,\ref{sect:mstar}) to derive metallicities of $12+\log({\rm O/H})=8.75\pm1.60$ at $z=0.40$, $8.65\pm0.71$ at $z=0.84$, $8.49\pm0.71$ at $z=1.47$ and $8.46\pm 0.65$ at $z=2.23$. 

Given these metallicities, the \citet{leroy11} relation (Equation\,\ref{eq:leroy}) implies gas-to-dust ratios $\delta_{\rm GDR}=90\pm30$, $110\pm20$, $150\pm20$ and $160\pm20$ for each of the four HiZELS stacks, the latter three of which are consistent with the typical Milky Way value $\delta_{\rm GDR}=140$ \citep{draine07}. In Appendix\,A\ref{app:fmr} we discuss the sensitivity of the gas masses estimated in this manner to the coefficients used to fit the FMR, comparing the results we derive using the \citet{stott13} fit for HiZELS galaxies to similar fits performed in [O{\sc ii}] and $r$-band selected samples in SDSS by \citet{mannucci10} and \citet{laralopez10}.

\subsection{The gas masses of HiZELS star-forming galaxies}\label{sec:gas-masses}

\begin{figure*}
\includegraphics[height=\textwidth,angle=90]{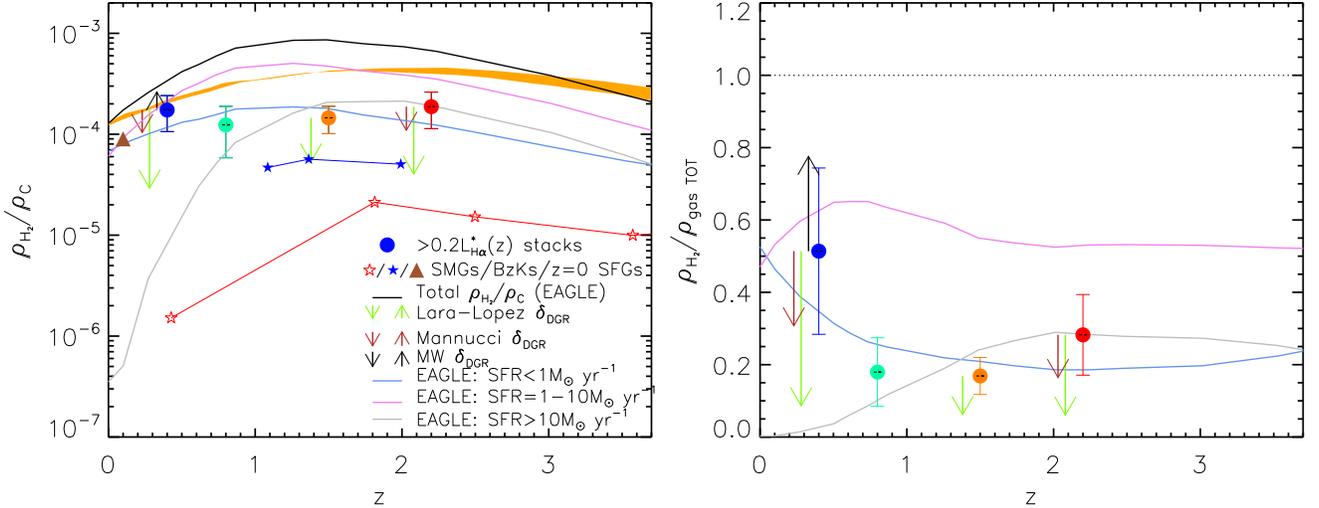}
\caption[Short captions]{\textit{Left:} The contribution of far-IR stacked $>0.2\,L^{\star}_{{\rm H}\alpha}$, star-forming galaxies to the total molecular gas (H$_2$) density of the universe from the {\sc eagle} simulation \citep[][thick black line]{lagos15}. Stacked H$\alpha$-selected galaxies are shown at $z=0.40, 0.84, 1.47$ and $2.23$ with filled, coloured points, using the same colour scale as Fig.~\ref{fig:seds}. Gas masses are measured from the dust continuum fits using the \citet{stott13} fundamental metallicity relation (FMR) to measure the metallicity from the SFR and $\langle M_\star\rangle$ of each stacked sample, and then using the relation between metallicity and the gas-to-dust ratio ($\delta_{\rm GDR}$) described by \citet{leroy11}. Coloured arrows to the left of each point show the effect of using the FMR fits of \citet{laralopez10} and \citet{mannucci10} to derive $\delta_{\rm GDR}$, or of using a fixed Milky Way $\delta_{\rm GDR}=140$ \citep{draine07}; we discuss the sensitivity of our results to the choice of FMR relation in detail in Appendix\,A\ref{app:fmr}. We do not plot arrows where the change is less than the $1\sigma$ error on the HiZELS data point. Also shown are the $z=0$ data from \citet{keres05}, along with higher-redshift $BzK$s and SMGs from \citet{daddi10a}, \citet{tacconi13} and \citet{swinbank14}, respectively, plus model predictions from \citet{lagos15} for the contribution of galaxies with given SFRs. We see that our stacked H$\alpha$-selected galaxies account for a significant fraction of the gas located in $1$--$10$\,M$_\odot$\,yr$^{-1}$ star-forming galaxies at $z=0.40$ and $2.23$. At $z=0.84$ and $1.47$, a significant fraction of the gas remains unaccounted for; however our analysis suggests that stacking H$\alpha$-selected SFGs recovers more of the gas in the universe than is measured by summing individually-detected $BzK$s/SMGs over the same redshift range. For comparison with {\sc eagle}, we also show the range of H$_2$ densities from the models of \citep[][orange band]{popping14}. \textit{Right:} As in the left panel, but normalised by the total gas density. We see that the our stacks all lie below the predicted contribution to $\rho_{\rm H_2}$ from galaxies with ${\rm SFR}=1$--$10$\,M$_\odot$\,yr$^{-1}$ in {\sc eagle}, indicating that the contribution made by HiZELS H$\alpha$-selected galaxies to the total gas density is significant, but that our H$\alpha$ cuts have also missed a significant contribution to the gas budget from low-SFR galaxies.}
\label{fig:madau}
\end{figure*}
\begin{table*}
\centering
\caption[Gas properties of HiZELS galaxies]{Gas properties of HiZELS galaxies}
\label{tab:gas-properties}
\begin{tabular}{lccccccccccc}
\hline
\multicolumn{1}{l}{} &
\multicolumn{1}{c}{$12+\log({\rm O/H})^a$} &
\multicolumn{1}{c}{$\delta_{\rm GDR}$} &
\multicolumn{1}{c}{$M_{\rm H_2}$} & 
\multicolumn{1}{c}{$\tau_{\rm dep}$} &
\multicolumn{1}{c}{$\tau_{\rm H}$} &
\multicolumn{1}{c}{$\rho_{\rm H_2}/\rho_{C}$} &
\multicolumn{1}{c}{$\rho_{\rm H_2}/\rho_{\rm H_2, TOT}$} &
\multicolumn{1}{c}{$f_{\rm gas}^b$} &\\
\multicolumn{1}{l}{} &
\multicolumn{1}{c}{} &
\multicolumn{1}{c}{} &
\multicolumn{1}{c}{($\times 10^{9}$\,M$_\odot$)} &
\multicolumn{1}{c}{(Gyr)} &
\multicolumn{1}{c}{(Gyr)} &
\multicolumn{1}{c}{(\%)} &
\multicolumn{1}{c}{(\%)} &
\multicolumn{1}{c}{} & \\
\hline
$z=0.40$  & $8.7\pm1.6$&$90\pm30$&$10.1\pm5.3$&$5\pm3$&$9.4$&$0.017\pm0.016$&$50\pm20$&$0.5\pm0.3$\\
$z=0.84$  & $8.7\pm0.7$&$110\pm20$&$3.5\pm2.0$&$0.5\pm0.3$&$6.6$&$0.012\pm0.007$&$20\pm10$&$0.7\pm0.4$\\
$z=1.47$  & $8.5\pm0.7$&$150\pm20$&$8.3\pm2.8$&$0.3\pm0.2$&$4.4$&$0.015\pm0.008$&$20\pm10$&$0.7\pm0.3$\\
$z=2.23$  & $8.5\pm0.6$&$160\pm20$&$4.6\pm1.9$&$0.1\pm0.1$&$3.0$&$0.019\pm0.007$&$30\pm10$&$0.5\pm0.3$\\
\hline
\end{tabular}
\\{\footnotesize $^a$We calculate $12+\log({\rm O/H})$ from $\langle {\rm SFR_{{\rm H}\alpha}}\rangle$ and $\langle M_\star \rangle$ using the fundamental metallicity relation (FMR) of \citet{stott13}, and use these metallicities to calclate $\delta_{\rm GDR}$ using the relation of \citet{leroy11}. We note that the \citet{leroy11} relation is sensitive to the metallicity, which in turn renders the results in this table sensitive to the choice of FMR (see Appendix\,A\ref{app:fmr} for details). $^b f_{\rm gas}$ includes both atomic and molecular components.}
\end{table*}
\vspace{5mm}

Using the \citet{stott13} fundamental metallicity relation in conjunction with the metallicity-dependent model of \citet{leroy11}, we estimate gas-to-dust ratios $\delta_{\rm GDR}\sim90$--$160$ for our H$\alpha$-selected sub-samples. Based on the atomic-to-molecular gas mass relation $\log_{10}(M_{\rm H_2})=0.99\times\log_{10}(M_{\rm HI})-0.42$, measured in a sample of local massive galaxies by \citet{saintonge11}, we infer a typical molecular-to-total gas mass ratio of $\sim 0.24$. The molecular gas masses implied by this analysis are summarised in Table\,\ref{tab:gas-properties}, and lie in the range ($3.5$--$10.1$)$\times10^{9}$\,M$_\odot$.  Hence the gas depletion timescales (assuming these galaxies maintain their current SFR) are $\tau_{\rm dep}=5\pm3$, $0.5\pm0.4$, $0.3\pm0.2$ and $0.1\pm0.1$\,Gyr for the galaxy samples at $z=0.40$, $0.84$, $1.47$ and $2.23$, respectively. 

We see from Table\,\ref{tab:gas-properties} that these molecular gas depletion timescales are significantly shorter than the Hubble time ($\tau_{\rm H}$) at each redshift (especially at $z\geq 0.84$). We perform a least-squares fit to the relation $\tau_{\rm dep}\sim(1+z)^{N}$, finding the data to be fit best by the exponent $N=-3.9\pm 1.0$. This exponent is steeper than that which is expected in a simple model where the gas depletion timescale is proportional to the dynamical timescale \citep[$N=-1.5$; ][]{dave11}. 

We measure total gas fractions ($f_{\rm gas}\equiv (M_\star + M_{\rm gas})/M_{\rm gas}$) of our H$\alpha$-selected stacks (including both atomic and molecular gas), finding $f_{\rm gas}=50\pm 30$\% at $z=0.40$, $70\pm 40$\% at $z=0.84$, $70\pm 30$\% at $1.47$ and $50\pm 30$\% at $z=2.23$. These gas fractions are high -- and carry significant uncertainties -- but are consistent with the results of \citet{tacconi13}, who measured the gas masses of star-forming galaxies at $z=1.2$ and $2.2$ via their \co\ \jthreetwo\ emission in the PHIBSS survey. If we assume an exponentially declining star formation history (SFH), then this (along with the short depletion timescales at $z>0.8$) suggests a scenario in which the typical star-forming galaxies near the peak of cosmic star formation activity -- which we select with our $>0.2L_{{\rm H}\alpha}^\star(z)$ cut --  must undergo replenishment of their gas reservoirs \citep[e.g.\,][]{tacconi13}. We note that while \citet{tacconi13} find evidence of decreasing gas fractions with increasing stellar mass, the large uncertainties in our gas mass estimates prevent us from identifying such a correlation within our sample.

It is illustrative to compare the contribution these H$\alpha$-selected galaxies make to the gas content of the Universe as a function of redshift, with predictions from models. For each of our four sub-samples of star formation rate selected galaxies, we estimate their contribution to the cosmological gas density (relative to the critical density at $z=0$, $\rho_{C\,z=0}\equiv 3H_{0}^{2}/(8\pi G)$) as

\begin{centering}
\begin{equation}\label{eq:hizels-rhog}
\frac{\rho_{\rm H_2}}{\rho_{C\,z=0}} = \frac{N_{\rm stack}\delta_{\rm GDR}M_{\rm dust}}{\rho_{C\,z=0} V_{{\rm H,}z}}
\end{equation}
\end{centering}

We find that the gas reservoirs of these H$\alpha$-selected star-forming galaxies account for $\sim1$--$2$\% of the critical density at each redshift.

In Fig.\,\ref{fig:madau}, we use our dust-based estimates of the molecular gas mass to measure the contribution made by H$\alpha$-selected star-forming galaxies to the total H$_2$ density of the Universe, as a function of redshift. We compare our gas mass densities with samples from the literature, including \citet{keres05}, who measured the gas content at $z=0$ as $\rho_{\rm gas}/\rho_{C\,z=0}\sim 1$\%, along with intermediate redshift observations of $z=0.2$--$0.8$ ULIRGs from \citet{combes11, combes13}, and $z\sim1$--$2.5$ BX/BM and $BzK$ star-forming galaxies from \citet{tacconi13} and \citet{daddi10a}. We also include the H$_2$ mass densities of $S_{870\,\mu{\rm m}}>1$\,mJy SMGs observed with ALMA from \citet{swinbank14}. 

We compare each of these populations with the latest cosmological hydrodynamical estimate of the evolving total H$_2$ gas budget of the Universe from the Evolution and Assembly of GaLaxies and their Environments \citep[{\sc eagle};][]{schaye15} simulation. Using the evolving total H$_2$ density from {\sc eagle} \citep{lagos15}, we find that H$\alpha$-selected galaxies brighter than $0.2L_{{\rm H}\alpha}^{\star}(z)$ account for $\sim 50\pm 20\%, 20\pm 10\%, 20\pm 10\%$ and $30\pm 10\%$ of the total cosmic gas supply at $z=0.40, 0.84, 1.47$ and $2.23$, respectively (see Table\,\ref{tab:gas-properties}). In comparison, the $BzK$ star-forming galaxies account for $\sim 10$\% of the total H$_2$ density from the EAGLE simulation at their redshifts, and the $\sim1$\,mJy SMGs typically less than 5\%. 

In their work with the {\sc eagle} simulation, \citet{lagos15} propose that the molecular hydrogen density, $\rho_{{\rm H}_2}$, is dominated at $z\lesssim 2$ by galaxies more massive than $\log_{10}[M_{\star}/{\rm M}_{\odot}]= 9.7$, at $z\gtrsim 2$ by galaxies less massive than $\log_{10}[M_{\star}/{\rm M}_{\odot}]= 9.7$, and at all redshifts, by galaxies with ${\rm SFR}\gtrsim 10$\,M$_\odot$\,yr$^{-1}$. Given the median $\langle {\rm SFR_{{\rm H}\alpha}}\rangle$ and $\langle M_\star \rangle$ of the galaxies in our sub-samples and the uncertainties in our gas mass estimates (which result in the large error bars in Fig.~\ref{fig:madau}), our results are broadly consistent with this picture; for $z>0.84$, as $\langle {\rm SFR_{{\rm H}\alpha}}\rangle$ and $\langle M_\star \rangle$ for our H$\alpha$-selected samples increase, so we find our galaxies contribute a (slightly) greater fraction of the total gas supply. This trend is exactly reversed with the $z=0.40$ measurement, however this may be the result of cosmic variance, given that the volume probed at this redshift is the smallest of any of the HiZELS redshift slices.

In order to further investigate the contribution made by H$\alpha$-selected galaxies to the evolving gas density of the Universe, we use estimates of $\rho_{\rm H_2}/\rho_{\rm C\,z=0}$ from the {\sc eagle} simulation, broken down as a function of SFR (Lagos, private communication), which allow us to obtain a prediction for the fraction of $\rho_{\rm H_2}/\rho_{\rm C\,z=0}$ held in \textit{all} galaxies whose star formation rate is higher than the lower limit implied by our $>0.2L^{\star}_{{\rm H}\alpha}(z)$ cut, regardless of whether those galaxies would be selected as narrow-band excess sources in the HiZELS survey or not. We find that H$\alpha$-selected galaxies brighter than $0.2L^{\star}_{{\rm H}\alpha}$ account for $100\pm30\%$ of the gas in galaxies at $z=0.40$ above our evolving SFR cut, falling to $40\pm 10\%$ at $z=0.84$, $20\pm 10\%$ at $z=1.47$, and $40\pm 10\%$ at $z=2.23$. Hence, we find that flux and equivalent width-limited H$\alpha$-selection provides an efficient means of selecting galaxies which host a significant fraction of the molecular gas in the Universe.

\section{Conclusions}\label{sec:conclusions}

In this paper, we have investigated the evolution of the far-IR properties of H$\alpha$-emitting star-forming galaxies across the peak of the star formation activity in the Universe. By bringing together $100$--$500\,\mu$m imaging from \herschel\ PACS/SPIRE, with $850\,\mu$m observations from SCUBA-2 and deep 1.4\,GHz VLA imaging, we have employed stacking in order to measure the far-IR properties of ``typical'' H$\alpha$-selected star-forming galaxies below the confusion limit, spanning the knee of the (evolving) H$\alpha$ luminosity function ($>0.2L^\star_{{\rm H}\alpha}(z)$). 

\begin{itemize}
\item We have measured the infrared luminosities, $L_{\rm IR}$, dust masses, $M_{\rm dust}$, and temperatures, $T_{\rm dust}$, of our stacked samples, finding them to evolve from cold ($\sim 14$\,{\sc k}) systems, with Milky Way-like luminosities ($L_{\rm IR}\sim 10^{10}$\,L$_\odot$) at $z=0.40$ to warmer, LIRG-like systems ($T_{\rm dust}\sim 34$\,{\sc k}, $L_{\rm IR}\sim 10^{11}$\,L$_\odot$) systems at $z=2.23$. The evolution in their far-IR properties is comparable to the evolution in $L_{{\rm H}\alpha}^\star(z)$ used to construct our samples.
\item Comparing the infrared and H$\alpha$-derived SFRs of those galaxies contributing to the stacks allows us to obtain new constraints on the levels of H$\alpha$ extinction due to dust. We see no evidence of strong trends in the ratio $\log_{10}[L_{{\rm H}\alpha}/L_{\rm IR}]$ with either redshift, $L_{\rm IR}$ or stellar continuum extinction, $A_V$, suggesting that our selection based on $L_{{\rm H}\alpha}$ relative to an evolving luminosity cut has selected galaxies with ``similar'' ISM properties at all redshifts.
\item Our stacking method allows us to study the relationship between the far-IR luminosities and dust temperatures of ``normal'' star-forming galaxies that are otherwise below the confusion limit in the \herschel\ SPIRE bands. Comparing the results of our stacking analysis with samples from the literature, we find our H$\alpha$-selected galaxies follow a similar trend in terms of $L_{\rm IR}$--$T_{\rm dust}$ to that seen in local galaxies, but are typically warmer (at a given $L_{\rm IR}$) than comparison samples at the same redshift. Using a modified Stefan-Boltzmann law, we estimate the characteristic sizes of the dust-emitting regions of HiZELS galaxies, finding them to be $\sim 0.5$\,kpc, nearly an order of magnitude smaller than their stellar sizes and providing tentative evidence of their having a clumpy ISM.
\item Using a two-step approach that entails estimating the dust metallicity from a ``fundamental metallicity relation'', and then calculating a metallicity-dependent gas-to-dust ratio, we use the far-IR photometry for our stacked H$\alpha$-selected samples to estimate their typical gas masses. By comparing these with the SFRs, we estimate gas depletion timescales $\tau_{\rm dep}\sim 0.1$--$5$\,Gyr across $z=2.2$--$0.40$. These short gas depletion timescales (relative to the Hubble time at each redshift) suggest that refuelling of these gas reservoirs -- likely either by steady accretion or by an accumulation of minor mergers -- may have taken place at some point in the past. The total gas mass fractions are high -- $f_{\rm g}\backsimeq 0.6\pm0.1$ -- at all redshifts. However, the choice of fundamental metallicity relation used to measure $\delta_{\rm GDR}$ is a potentially significant systematic, and accounts for an additional factor $\sim2\times$ uncertainty on the gas masses at all redshifts (Appendix\,A\ref{app:fmr}).
\item Combined with the number densities of H$\alpha$-selected galaxies at $z=0.40$, $0.84$, $1.47$ and $2.23$, this allows us to estimate the contribution of $>0.2L_{{\rm H}\alpha}(z)$, H$\alpha$-selected galaxies to the evolving H$_2$ content of the Universe. We find that galaxies satisfying our selection criteria comprise a significant fraction of the total H$_2$ in the Universe ($35\pm10$\% on average), as predicted by the state-of-the-art cosmological hydrodynamical simulation {\sc eagle}, modulo the aforementioned uncertainties due to the FMR.
\end{itemize}

Our results are important in order to understand the nature and evolution of luminous H$\alpha$ emitters within the context of the evolving properties of the star-forming galaxy population. As the typical of SFRs of galaxies increase from the local Universe towards the peak of star formation at $z=1$--$2$, so too does the relative prevalence of dusty systems, observable in the far-IR, leading to increasing biases in surveys that rely exclusively on UV/optical tracers at higher redshifts.

\section*{Acknowledgements}
The authors wish to thank the anonymous reviewer for helpful comments which improved the quality of this manuscript. APT, IRS and AMS acknowledge support from STFC (ST/L00075/X). APT and IRS also acknowledge support from the ERC Advanced Investigator Programme {\sc dustygal} (\#321334); IRS acknowledges a Royal Society/Wolfson merit award. DS acknowledges financial support from the Netherlands Organisation for Scientific Research (NWO) through a Veni Fellowship. DS also acknowledges funding from FCT through a FCT Investigator Starting Grant and Start-up Grant (IF/01154/2012/CP0189/CT0010) and from FCT grant PEst-OE/FIS/UI2751/2014. APT thanks John Stott and Claudia Lagos for sending insightful comments on the fundamental metallicity relation, and sharing tables of data from {\sc eagle}, respectively. The SCUBA-2 $850\,\mu$m data presented in this paper were taken as part of Program ID MJLSC02. It is a pleasure to thank the entire staff of the JCMT for their superb support throughout the S2CLS campaign. The National Radio Astronomy Observatory is a facility of the National Science Foundation operated under cooperative agreement by Associated Universities, Inc. This research has made use of data from HerMES project (\url{http://hermes.sussex.ac.uk/}). HerMES is a \herschel\ Key Programme utilising Guaranteed Time from the SPIRE instrument team, ESAC scientists and a mission scientist. \textit{Herschel}-ATLAS is a project with \herschel, which is an ESA Space Observatory with science instruments provided by European-led Principal Investigator consortia and with important participation from NASA. The H-ATLAS Web site is \url{www.h-atlas.org}. The authors wish to thank all staff at the (former) JAC and their successors at EAO for their help in conducting the observations at the UKIRT telescope and their ongoing support.

\clearpage

\appendix
\subsection{A.1 -- The effect of clustering on stacked flux densities}\label{app:clustering}
We investigate possible biases in our stacking analysis due to the inherent clustering of galaxies within the large-beam SPIRE images by performing a set of simulations, using model images of the COSMOS field at 250, 350 and 500\,$\mu$m.

We begin with a blank grid, using the same astrometry and pixel scale as for the real SPIRE images, and add delta functions at the positions of each of the $\sim 47,000$ $24\,\mu$m/radio-detected galaxies in the prior catalogue, whose flux densities in each of the model images are set to the flux densities measured when the raw SPIRE images were originally deblended. Next, we seed the maps with delta functions at the positions of the $1771$ HiZELS galaxies in COSMOS (applying a $0.5''$ exclusion radius around each galaxy to exclude those HiZELS galaxies that were also priors, in order to avoid injecting the same galaxies twice). The SPIRE flux densities of these model HiZELS galaxies in the model SPIRE images are set to those of a $30\,${\sc k} greybody SED (the weighted average temperature of the real HiZELS SEDs) at the appropriate redshifts, scaled by each galaxy's measured $L_{{\rm H}\alpha}$ under the condition that ${\rm SFR}_{\rm IR} = {\rm SFR}_{{\rm H}\alpha}$. We convolve each grid of delta functions with the appropriate SPIRE PSF, before adding Gaussian random noise from a distribution matching the rms of the original residual images (i.e.\,$1.6$\,mJy, $2.1$\,mJy and $3.1$\,mJy at $250$, $350$ and $500\,\mu$m, respectively). We then deblend these model SPIRE images using the same technique as was used for the real maps, requiring the code to fit sources in each of the bands at the positions of every galaxy injected in to the image (i.e.\,$24\,\mu$m/radio priors \textit{and} model HiZELS galaxies). This generates a new list of deblended photometry, along with a new residual image at each wavelength. We perform the same stacking analysis as outlined in \S\,\ref{sect:spire-deblending}, by reinjecting the model HiZELS sources in to the model residual images and stacking those galaxies that lie above the evolving cut, $L_{{\rm H}\alpha}\geq 0.2L^{\star}_{{\rm H}\alpha}$ (these simulations are called ``clustered HiZELS stacks'').

We then repeat this entire process, generating new model maps by injecting scaled delta functions at the positions of the $24\,\mu$m/radio-detected galaxies; however, this time, we inject the model HiZELS greybodies at random positions in the field, and run the deblending code again. By randomising the positions of the model HiZELS galaxies, we erase their clustering signature. We again stack the $L_{{\rm H}\alpha}\geq 0.2L^{\star}_{{\rm H}\alpha}$ model HiZELS galaxies by reinjecting sources in to the residual images at these randomised positions, and measure their $250$, $350$ and $500\,\mu$m flux densities (``unclustered HiZELS stacks'').

By comparing the flux densities measured from these two sets of simulated SPIRE stacks, it is possible to directly quantify the effect the clustering of HiZELS galaxies with each other has on their measured stacked flux densities. We find that the differences between the flux densities of our ``clustered'' and ``unclustered'' model HiZELS galaxies are ($0.05\pm 0.03$)\,mJy at $250\,\mu$m, ($-0.02\pm 0.04$)\,mJy at $350\,\mu$m and ($0.07\pm 0.09$)\,mJy at $500\,\mu$m. We therefore find no evidence that the clustering of HiZELS galaxies (with each other) has significantly boosted the flux densities of our stacks (presented in Table\,\ref{tab:hizels-photometry}).

A final source of potential bias is the clustering of (non-H$\alpha$ detected) field galaxies around the HiZELS galaxies, i.e.\ galaxies which contribute flux in the real SPIRE images but are not included in either the deblending prior list or our target list. Since the deblending code is required to minimise the residuals (i.e.\,$\mid data - model \mid$), and since the typical source density of optically-selected galaxies per SPIRE beam is high, the effect of having a galaxy population in the SPIRE images that is not in the prior list would be that their flux would be erroneously ``pulled in'' and assigned to the positions of the priors. If those prior galaxies were then reinjected (with their best-fit fluxes) into the residual images (following the process used for the real data) and stacked, their stacked flux densities would be higher than their ``true'' flux densities. We attempt to quantify this effect by repeating the deblending process yet again, generating a new set of model SPIRE images comprising: (i) the $47000$ $24\,\mu$m/radio priors; (ii) the $1771$ model HiZELS galaxies; and (iii) $295,000$ field galaxies at the positions of $I$-band selected galaxies in COSMOS \citep[][again with a $0.5''$ exclusion radius around each prior/HiZELS galaxy to avoid injecting the same galaxies multiple times. This step excludes $\sim 45000$ duplicates from the $I$-band catalogue.]{ilbert08}. We inject these field galaxies into the SPIRE images with fluxes set to those of greybody SEDs with $T_{\rm dust}$ drawn randomly in the range $25$--$50$\,{\sc k}, and shifted to the photometric redshifts estimated by \citet{ilbert08}. Each of these field galaxy SEDs is scaled in luminosity to an SFR drawn randomly between ${\rm SFR}=0.1$--$10$\,M$_\odot$\,yr$^{-1}$. The median flux densities of these field galaxies are $S_{250}=(0.756\pm0.012)$\,mJy, $S_{350}=(0.405\pm 0.005)$\,mJy and $S_{500}=(0.175\pm0.002)$\,mJy. We deblend these new model images \textit{using the same prior list as before}, and once again stack at the positions of the $>0.2L^\star_{{\rm H}\alpha}(z)$ galaxies (these are termed ``stacks with field galaxies''). By comparing the flux densities measured in the ``with field galaxies'' and ``clustered HiZELS'' stacks, we are able to quantify the degree to which the fluxes of our HiZELS stacks may have been boosted due to the presence of a population of clustered field galaxies, which isn't accounted for in the deblending.

We find offsets in the flux densities of our ``with field galaxies'' and ``clustered HiZELS'' stacks to be $\Delta S_{\rm 250}=-0.5\pm 0.2$\,mJy, $\Delta S_{\rm 350}=0.2\pm 0.3$\,mJy, $\Delta S_{\rm 500}=0.2\pm 0.5$\,mJy. These offsets are comparable to the statistical uncertainties in the stacked flux densities from the real images at each wavelength and are not systematically dependent on wavelength/beam size, therefore we do not apply a correction to the flux densities measured from stacking in the real SPIRE data. Instead, we incorporate this effect in to our results by combining the size of the offsets in flux densities between our simulations and the statistical uncertainties in our stacks in quadrature (Table\,\ref{tab:hizels-photometry}). From this series of tests, we conclude that the contribution to the flux densities in the SPIRE images from field galaxies \textit{not} in our prior catalogue is low (at least in the regions around the SFR-selected HiZELS galaxies). We note that the magnitude of this effect is likely to be a function of the flux distribution of the field galaxy population, which is the very unknown that stacking analyses are intended to address. In our simulations, galaxies are injected in to the model SPIRE images with flux densities that are appropriate given the \textit{reasonable}, physically-motivated assumptions we have made for their observed-frame colours and SFR. In a forthcoming paper, Cochrane \etal\ (submitted) will show that HiZELS galaxies are typically star-forming centrals located in relatively low mass haloes ($M_{\rm halo}\sim 10^{12}$--$10^{13}$\,M$_\odot$), which may provide a physical explanation for our finding that contamination in the far-IR from galaxies \textit{other} than those we have already accounted for in the deblending process is low.

\subsection{A.2 -- Sensitivity of metallicities to the SFR and stellar mass}\label{app:fmr}

In \S\,\ref{sec:madau}, we measured the gas masses of our HiZELS galaxies using a two-step process in which we used dust masses measured from our far-IR SED fits, and converted these to total (H{\sc i}+H$_2$) gas masses using a metallicity-dependent gas-to-dust ratio, $\delta_{\rm GDR}$. Because we do not have emission line diagnostics to constrain the metallicity for each of our HiZELS galaxies, we exploited the mass-metallicity-SFR relation (the so-called``fundamental metallicity relation''; Equation\,\ref{eq:fundamentalplane}) to infer the typical metallicities of our stacked sub-samples from their broad-band photometric properties. Our analysis utilised the parameterisation of the FMR carried out by \citet{stott13}, who measured its shape in a sub-set of the H$\alpha$-selected galaxies present in this sample, finding the metallicities of HiZELS galaxies to be a strong function of the SFR, and with only a weak dependence on the product of the stellar mass and SFR entering via the $a_4$ coefficient.

A number of works have sought to characterise the FMR, using photometry and spectral line data drawn from different samples of galaxies. Notably, \citet{mannucci10} studied the FMR in a sample of $z\sim 0.1$, [O{\sc ii}]-selected star-forming galaxies from the Sloan Digital Sky Survey (SDSS), finding it to be well-characterised by the coefficients $a_0=8.90$, $a_1=0.37$, $a_2=-0.14$, $a_3=-0.19$, $a_4=0.12$ and $a_5=-0.054$, while in another work, \citet{laralopez10} found the surface of the FMR for $z\sim 0.1$, $r$-band selected galaxies in SDSS-DR7 to be fit by $\log(M_\star)= 1.122\times[12+\log({\rm O/H})] + 0.474\times\log ({\rm SFR}) - 0.097$.

If we use the \citet{mannucci10} and \citet{laralopez10} measurements of the FMR in lieu of the \citet{stott13} FMR, respectively, then the metallicities of our samples are raised to $[12+\log({\rm O}/{\rm H}]=9.01\pm0.75$ ($9.66\pm 1.71$) at $z=0.40$, $8.72\pm0.48$ ($8.99\pm1.91$) at $z=0.84$, $8.63\pm0.54$ ($9.01\pm3.22$) at $z=1.47$ and $8.73\pm0.53$ ($9.24\pm3.32$) at $z=2.23$. Following Equation\,\ref{eq:leroy}, these metallicities imply gas-to-dust ratios $\delta_{\rm GDR}=55\pm9$ ($15\pm5$), $96\pm10$ ($57\pm23$), $114\pm14$ ($55\pm$) and $94\pm11$ ($34\pm24$), at the four HiZELS redshifts respectively.

We see that the higher metallicities implied by the \citet{laralopez10} FMR fit act to lower $\delta_{\rm GDR}$ considerably ($\sim3$--$6\times$) compared to the \citet{stott13} fit, whereas the \citet{mannucci10} FMR and \citet{stott13} fits typically agree with each other to within a factor $\sim 2\times$. While a quantitative analysis of the fundamental metallicity relation is beyond the scope of this paper, it is important to bear in mind the implications these three measurements have for our gas mass measurements in \S\,\ref{sec:madau}. To begin to understand the reasons why these three measurements of the FMR differ, it is important to understand the compositions of the samples in which they were measured. The galaxies we stack in our sub-samples are H$\alpha$-selected, with $>0.2L^\star_{{\rm H}\alpha}(z)$, and typical $\langle{\rm SFR}\rangle = 20$\,M$_\odot$\,yr$^{-1}$ and $\langle \log M_\star \rangle = 10.5$ (see Table\,\ref{tab:stack-properties}). The analysis of \citet{stott13} is based on a (different) subset of galaxies drawn from the same (HiZELS) parent sample with similar mass ($\langle \log M_\star \rangle = 9.5$), albeit with slightly lower typical SFR ($\langle{\rm SFR}\rangle =11$\,M$_\odot$\,yr$^{-1}$). The works of \citet{mannucci10} and \citet{laralopez10} are both based on significantly less active galaxies ($\langle{\rm SFR}\rangle=1.5$\,M$_\odot$\,yr$^{-1}$) at lower-redshift, albeit at a similar stellar mass ($\langle \log M_\star \rangle=10.1$).

Recent work on the nature of the FMR by \citet{telford16}, using updated line ratio diagnostics, finds a generally weaker anti-correlation between metallicity and SFR (at a given $M_\star$) than that reported in \citet{laralopez10}, which may reconcile some of the $\sim3$--$6\times$ discrepancy between these $\delta_{\rm GDR}$ estimates. However, this anti-correlation is thought to be stronger in galaxies whose current or recent SFRs are higher than their past average SFRs. Without knowing the detailed star formation histories of our H$\alpha$-selected galaxies, or having the line ratio diagnostics necessary to directly measure the FMR in our H$\alpha$-selected sub-samples, we have little choice but to adopt an FMR; we chose to adopt the \citet{stott13} FMR, as it is calculated in the region of SFR-$M_\star$ parameter space that most closely matches that of our $>0.2L^\star_{{\rm H}\alpha}$ sub-samples, but we reiterate that the gas-to-dust ratios (and hence gas masses) we subsequently derive are sensitive to this choice of FMR.

\bibliographystyle{mn2e} 
\bibliography{ref.bib}

\end{document}